\documentclass[11pt]{article}
\usepackage{graphicx}
\usepackage{epsfig}
\usepackage{amsfonts}
\usepackage{amssymb}
\usepackage[dvips]{color}

\newcommand{\insertplot}[5]{\begin{figure}
 \hfill\hbox to 0.05in{\vbox to #5in{\vfill
 \inputplot{#1}{#4}{#5}}\hfill}
 \hfill\vspace{-.1in}
 \caption{#2}\label{#3}
 \end{figure}}
 \newcommand{\inputplot}[3]{
 \special{ps: plotfile #1}
}
\newcounter{fig}

\newcommand{\ee}{\end{equation}}
\newcommand{\eea}{\end{eqnarray}}
\newcommand{\be}{\begin{equation}}
\newcommand{\bea}{\begin{eqnarray}}

\date{}

\begin{document}

 \title{Magnetic and Electric Black Holes \\
 in the Vector-Tensor Horndeski Theory }

\author{
{\large Y. Verbin}
\\
\\
{\small Astrophysics Research Center, the Open University of Israel, Raanana 4353701, Israel}
\\
}

\maketitle

\begin{abstract}
We construct exact solutions of magnetically charged black holes in the vector-tensor Horndeski gravity
and discuss their main features. Unlike the analogous electric case, the field equations are linear in a simple (quite standard) parametrization of the metric tensor and they can be solved analytically even when a cosmological constant is added. The solutions are presented in terms of hypergeometric functions which makes the analysis of the black hole properties relatively straightforward. Some of the aspects of these black holes are quite ordinary like the existence of extremal configurations with maximal magnetic charge for a given mass, or the existence of a mass with maximal temperature for a given charge, but others are somewhat unexpected, like the existence of black holes with a repulsive gravitational field. We perform our analysis for both signs of the non-minimal coupling constant and find black hole solutions in both cases but with  significant differences between them. The most prominent difference is the fact that the black holes for the negative coupling constant have a spherical surface of curvature singularity rather than a single point. On the other hand, the gravitational field produced around this kind of black holes is always attractive. Also, for small enough magnetic charge and negative coupling constant, extremal black holes do not exist and all magnetic black holes have a single horizon. In addition we study the trajectories around these magnetic black holes for light as well as massive particles either neutral or electrically charged. Finally, we compare the main features of these black holes with their electric counterparts, adding some aspects that have not been discussed before, like temperature,  particle trajectories and light deflection by electrically charged Horndesky black holes.
 \end{abstract}

\section{Introduction}
	
    In the effort aimed to understand the dark matter and dark energy
problems of the Universe, numerous extensions of General Relativity (GR) have been
studied in the last few decades. Among these generalized gravities the extensions of the
minimal Einstein-Hilbert Lagrangian by scalar fields play an important role.
With the main objective of maintaining field equations of the second order in field derivatives, thus avoiding the Ostrogradsky instability, the works of G. Horndeski \cite{Horndeski:1974wa} provoked a considerable revival of interest in the last years with a vast range of applications from cosmology to black hole (BH) physics.

A family of vector-tensor theories was also found by Horndeski \cite{Horndeski:1976gi} as an answer to the analogous question: what is the most general extension of the Einstein-Hilbert Lagrangian by a vector field which analogously keeps the field equation of second order with the additional conditions that gauge invariance is still valid and such that the electromagnetic equations reduce to Maxwell's equation in the absence of gravity. Unlike the scalar-tensor Horndeski theory which has a huge freedom, the vector-tensor theory is essentially unique. It is characterized by a single interaction term (which we call Horndeski term) between the geometry and the vector field with a single coupling constant. The action considered is of the form
\be
   S = \int d^4 x \sqrt{-g} \bigg[ \frac{1}{2\kappa} R	- \frac{1}{4} F_{\mu \nu}F^{\mu \nu}  -\frac{\gamma \kappa}{4} (F_{\mu \nu} F^{\kappa \lambda} R^{\mu \nu}_{\phantom{\mu \nu} \kappa \lambda} - 4 F_{\mu \kappa} F^{\nu \kappa} R^{\mu}_{{\phantom \mu} \nu}
+ F_{\mu \nu} F^{\mu \nu} R )  \bigg]
\label{Lagrangian}
\ee
where $F_{\mu \nu}$ is the electromagnetic field strength and $R^{\mu \nu}_{\phantom{\mu \nu} \kappa \lambda}$ is the Riemann tensor with $R^{\mu \nu}_{\phantom{\mu \nu} \mu \lambda}=R^\nu_{\;\lambda}$ and $R^\nu_{\;\nu}=R$.  We use $\kappa = 8 \pi G $ which we later take to be 1 by rescaling. $\gamma$ is a dimensionless parameter which fixes the strength of the Horndeski non-minimal coupling. The last term ${\cal I}(g,A)$ is the non-minimal coupling term of the vector field to the geometry introduced by Horndeski \cite{Horndeski:1976gi} as the only possible interaction term which still keeps the field equations of second order:
\be
\label{HorndeskiTerm}
{\cal I}(g,A) = \frac{1}{4}\; ^{**}R^{\mu \nu}_{\phantom{\mu \nu} \kappa \lambda} F^{\kappa \lambda}F_{\mu \nu}= -\frac{1}{4} (F_{\mu \nu} F^{\kappa \lambda} R^{\mu \nu}_{\phantom{\mu \nu} \kappa \lambda} - 4 F_{\mu \kappa} F^{\nu \kappa} R^{\mu}_{{\phantom \mu} \nu}
+ F_{\mu \nu} F^{\mu \nu} R )
\ee
where $^{**}R^{\mu \nu}_{\phantom{\mu \nu} \kappa \lambda}$ is the doubly dual Riemann tensor. Similarly, we use the dual field strength $^{*}F^{\mu \nu}$. Both are defined by applying appropriately the Levi-Civita tensor $\sqrt{-g}\epsilon_{\kappa \lambda \mu \nu} $.

There are various ways to write down the field equations derived from (\ref{Lagrangian}). We use the following form:
\bea
\label{FEqA}
\nabla_\mu \left(F^{\mu\nu} - \gamma\kappa\;\; ^{**}R^{\mu \nu}_{\phantom{\mu \nu} \kappa \lambda} F^{\kappa \lambda} \right)=0\\\label{FEqg}
G_{\mu\nu}+\kappa T^{(Max)}_{\mu\nu} - \gamma\kappa^2 H_{\mu\nu}=0
\eea
where $T^{(Max)}_{\mu\nu}$ is the Maxwell standard contribution of the energy-momentum tensor and the contribution from the Horndeski term is:
\bea
\label{H-Tensor} \nonumber
H_{\mu\nu} = \frac{1}{4}g_{\mu\nu} \;  ^{**}R_{ \kappa \lambda}^{\phantom{\mu \nu} \rho\sigma} F^{\kappa \lambda}F_{\rho\sigma} - \frac{1}{2}\left( ^{**}R_{\lambda\mu}^{\phantom{\mu \nu} \rho\sigma} F^{\lambda}_{\phantom{\mu} \nu}+ \;
^{**}R_{\lambda\nu}^{\phantom{\mu \nu} \rho\sigma} F^{\lambda}_{\phantom{\mu} \mu}  \right)F_{\rho\sigma}  \\
-R^{\rho\sigma} \;^{*}F_{\rho\mu} \; ^{*}F_{\sigma \nu}+(\nabla_\kappa \;^{*}F_{\mu}^{\phantom{\mu}\lambda})(\nabla_\lambda \;^{*}F_{\nu}^{\phantom{\mu}\kappa})
\eea

In a sharp distinction with respect to the scalar-tensor Horndeski theory, relatively little effort was invested in its vector-tensor relative. The first studies of this theory were naturally done in the static spherically-symmetric case, first by Horndeski himself \cite{Horndeski:1978ca} and then in more detail by Muller-Hoissen and  Sippel  \cite{muller} who found that the electric solutions contain deformations of the Reissner-Nordstrom (RN) solutions, which may be described as Horndeski-Reissner-Nordstrom (HRN) electrically charged black holes. Some additional work clarifying open points followed several years later \cite{BalakinEtAl2008}. Further work was done recently in the context of a scalarized version of these solutions \cite{YB+YV2020}.

In a parallel path, the cosmological aspects of this vector-tensor theory were studied by several authors \cite{EspositoFareseEtAl2009,Barrow:2012ay,BeltranJimenezEtAl2013}  and others considered also the non-Abelian version \cite{Davydov:2015epx,BeltranJimenezEtAl2017}.

The Horndeski non-minimal term (\ref{HorndeskiTerm}) appears also, alongside with several other coupling terms,  in further generalized vector-tensor theories \cite{Heisenberg2014,Tasinato2014} that break gauge invariance such as generalized Proca theories \cite{HeisenbergEtAl2017a}. Cosmological solutions \cite{Tasinato2014} as well as spherically-symmetric  solutions and electric BHs \cite{HeisenbergEtAl2017a,HeisenbergEtAl2017b} of these theories were constructed. Some of the BH solutions were obtained in a closed analytic form. In this context it is useful to note that although the generalized vector-tensor theories which break gauge invariance apparently contain the vector-tensor Horndeski theory, they cannot produce in a certain limit (like the limit of vanishing mass of the generalized Proca theories) the solutions of the gauge invariant  vector-tensor Horndeski theory.

In this paper we return to the localized static spherically-symmetric  solutions, but now \emph{magnetically} charged. Only very little exists in the literature about the magnetic counterpart of the electrostatic non-minimal BHs mentioned above. Perhaps the reason is the experience from the pure Einstein-Maxwell system where the magnetic black hole is essentially identical to the Reissner-Nordstrom solution, although it is obvious that this should not be the case since the duality symmetry is broken by the Horndeski term ${\cal I}(g,A)$.

 The first work about magnetically charged black holes with the Horndeski non-minimal coupling (MHBH for short) was a short study in a rather unknown paper by Horndeski \cite{Horndeski-Birkhof+Mag1978} that has accumulated 11 citations todate. In addition there are several more recent works \cite{BalakinZayats2007,BalakinEtAl2016A} which concentrate  on the non-Abelian generalization of the Horndeski vector-tensor theory usually containing a larger family of non-minimal coupling terms which yield field equations of order higher than 2. These papers present self-gravitating magnetic monopoles of the Wu-Yang type and magnetic BHs with further extensions like adding a cosmological constant. Some exact solutions have been found too \cite{BalakinEtAl2016B}, but they are solutions to some special cases which do not include the Horndeski coupling.

Here we revisit the Abelian theory and show that it deserves  further study. A significant part of the study here is based on the finding that the field equations for the magnetically charged Horndeski BHs can be casted as two decoupled linear differential equations that can be solved analytically.

After completion of this work, I became aware of Ref. \cite{Feng+Lu2015} which constructs and studies in the context of the AdS/CFT correspondence,  a general family of vector-tensor theories with non-minimal coupling terms in arbitrary number of dimensions which are still gauge-invariant and produce second order field equations. These theories were further extended to include also $p$-form gauge field strengths. Black hole solutions of this large family of theories were studied assuming from the outset the presence of a cosmological constant and extending the horizon topologies beyond 2-sphere, to a 2-torus and hyperbolic 2-space.
Additional papers followed this route of general gauge invariant higher dimensional higher curvature theories to various directions like providing consistent Kaluza-Klein compactifications of Lovelock gravity using magnetic monopole configurations to dress the internal manifold of spacetime \cite{CisternaEtAl2020A} or construct black hole solutions and regular multi-horizon black holes \cite{CisternaEtAl2020B}. The family of these theories has the Horndeski vector-tensor theory of Eq. (\ref{Lagrangian}) as a special case and the results of Ref. \cite{Feng+Lu2015} have some overlap  with the present paper, in particular the explicit exact solution of the Magnetic BHs presented in the next sections and some of their properties. These solutions appear in a different context and representation in sec. 4.3 of Ref. \cite{Feng+Lu2015}.

After presenting the MHBH solutions in Sec. \ref{sphsymmsol}, we discuss in Sec. \ref{CharactreisticsSolGamPos} their main general characteristics with respect to the ordinary RN solutions, like horizon pattern, the relation among the BH mass, charge and horizon, the temperature and so on for $\gamma>0$, and then in Sec.  \ref{CharactreisticsSolGamNeg} for $\gamma<0$. The geometrical structure of these MHBHs spacetimes will be analyzed in Secs. \ref{GeodesicsNeut} and \ref{LightDeflection}
using timelike and null geodesics. Light deflection will be studied in detail in Sec. \ref{LightDeflection}.
In addition, the trajectories of charged particles will be described briefly in Sec. \ref{GeodesicsCharged}.

Although the basics of the electric counterparts of the MHBHs are known for a long time \cite{muller}, there still exists a gap to be  filled in analogy to the studies of the above mentioned sections. Sec. \ref{ElectricBHs} will be therefore dedicated to first obtaining (numerically) the electric BHs and then presenting their other characteristics and comparing to the magnetic type of solutions. In Sec. \ref{LightDeflectionEl} we will present particle trajectories and light deflection. Sec. \ref{conclusion} will contain the conclusion.

\section{The model: Magnetic Spherically-Symmetric Solutions}\label{sphsymmsol}
\setcounter{equation}{0}
We are interested in magnetic spherically-symmetric solutions for the
Einstein-Maxwell-Horndeski field equations (\ref{FEqA})-(\ref{FEqg}) with (\ref{H-Tensor}).

\subsection{Ansatz and Field Equations}\label{ansatz}
In order to obtain static spherically symmetric solutions we will adopt  a very popular parametrization of the metric
\be
     ds^2 = f(r) a^2(r) dt^2 - \frac{1}{f(r)} dr^2 - r^2 d \Omega_2^2
\label{LineElement}
\ee
completed by a spherically-symmetric magnetic field derived from the vector potential $A_\mu dx^\mu=P(1-\cos \theta)d\phi$. The magnetic function $P$ must be constant since we insist on spherical symmetry. In that case $P$ is just the magnetic charge. Without loss of generality we assume $P>0$. Incidentally, we note that there can be no magnetic charge in the analogous spherically-symmetric solutions of the generalized vector-tensor theories mentioned above \cite{HeisenbergEtAl2017a,HeisenbergEtAl2017b}.

Eq. (\ref{FEqA}) is thus satisfied trivially, but from Eq (\ref{FEqg}), or (what is easier), using directly the Lagrangian $\sqrt{-g}\cal{L}$, one finds after some elementary manipulations the following 2 decoupled \emph{linear} equations:
\be
\label{Eqa}
 (r^4+\gamma\kappa^2 P^2)r\frac{a'}{a}  + 3\gamma\kappa^2 P^2 = 0
\ee
\be
\label{Eqf}
 (r^4+\gamma\kappa^2 P^2)rf'   +(r^4-6\gamma\kappa^2 P^2)f+  \frac{\kappa P^2}{2}r^2-r^4=0
\ee
There exists a third (second order) equation which is not independent and we do not present here.

\noindent Occasionally, we will use also the accumulated mass function $M(r)$ defined by $f(r)=1-2M(r)/r$.

\subsection{Solutions of the Field Equations}
The equation for the function $a(r)$ is easily solved by:
 \be
\label{Sola}
a(r) =\left | 1+ \frac{\gamma\kappa^2 P^2}{r^4} \right |^{3/4}
\ee
where the integration constant is taken such that $a(r)\rightarrow 1$ asymptotically. The absolute value is added in order to take care of the case $\gamma<0$. The second equation is less trivial to solve, and it is simpler to distinguish between two cases: $\gamma>0$ and $\gamma<0$.
\subsubsection {$f(r)$ for $\gamma>0$}
For $\gamma>0$ we change variables such that $z=\gamma\kappa^2 P^2/r^4$ and get for $f(z)$ the following linear and quite simple equation:
\be
\label{Eqfz}
 4(z+1)z f' +(6z-1)f  -  p\, z^{1/2} + 1=0
\ee
where $p=P/2\gamma^{1/2}$. Notice that this equation contains a single free parameter, $p$ which will be one of the characteristics of the BH solutions. A second one will be an integration constant which will determine their mass.

The solution of this equation can be written explicitly and analytically in terms of the Gauss hypergeometric functions $F(a,b,c,z)$ as (see Appendix):
\be
\label{Solfz}
f(z)=\frac{1}{ (1+z)^{7/4}} \left[-\mu z^{1/4}+p z^{1/2}F\left(-\frac{3}{4},\frac{1}{4},\frac{5}{4},-z\right) +F\left(-\frac{3}{4},-\frac{1}{4},\frac{3}{4},-z\right) \right]
\ee
The solution is parametrized by the integration constant $\mu$ which is obviously related to the mass as we see shortly. The dependence on the non-minimal coupling constant $\gamma$ is actually absorbed in the dimensionless parameters $\mu$ and $p$. In terms of the  dimensionless radial coordinate $x=r/(\gamma^{1/4}\sqrt{\kappa P})= z^{-1/4}$ the solution reads:
 \bea
\label{Solfx}
f(x)=\frac{1}{ (1+1/x^4)^{7/4}} \left[-\frac{\mu }{x}+\frac{p}{x^2} F\left(-\frac{3}{4},\frac{1}{4},\frac{5}{4},-\frac{1}{x^4}\right) +F\left(-\frac{3}{4},-\frac{1}{4},\frac{3}{4},-\frac{1}{x^4}\right) \right]
\eea
and the mass of the MHBH will be obtained from the asymptotic behavior of $f(r)$:
\be
\label{Asymptf}
f(r)=1-\frac{2M}{r}+\frac{\kappa P^2}{2 r^2}-\frac{2\gamma\kappa^2 P^2}{ r^4}+\frac{7\gamma\kappa^2 P^2 M}{2 r^5}-\frac{4\gamma\kappa^3 P^4}{5 r^6}+...
\ee
that is, the coefficient of the $1/r$ term is related to the integration constant such that $2M = \mu\gamma^{1/4}\sqrt{\kappa P}$. The explicit form of $f(r)$ is obtained trivially from (\ref{Solfx}):
  \bea
\label{Solfr}
f(r)=\left(1+ \frac{\gamma\kappa^2 P^2}{r^4}\right)^{-7/4} \left[ F\left(-\frac{3}{4},-\frac{1}{4},\frac{3}{4},-\frac{\gamma\kappa^2 P^2}{r^4}\right)-\frac{2M}{r}+\frac{\kappa P^2}{2 r^2} F\left(-\frac{3}{4},\frac{1}{4},\frac{5}{4},-\frac{\gamma\kappa^2 P^2}{r^4}\right) \right]
\eea

 As in the electric case, it depends on the two ``hairs'', mass and magnetic charge. Note also that taking $\gamma=0$ in the solution, it goes over to the magnetic Reissner-Nordstrom solution, i.e. the first 3 terms in the asymptotic expansion above. For $P=0$ the solution reduces of course to Schwarzschild. Schwarzschild (\textbf{S}) solution is also a solution of the full system ($\gamma\neq 0$) if there is no magnetic charge, while the RN solution is not a solution in any circumstances. Actually, the solution (\ref{Solfr}) is just a modification of the RN solution by the overall prefactor $(1+\gamma\kappa^2 P^2/r^4)^{-7/4}$ and the hypergeometric functions which multiply the RN terms.


\subsubsection {$f(r)$ for $\gamma<0$}

For $\gamma<0$  one defines $z=|\gamma|\kappa^2 P^2/r^4$ and $p=P/2|\gamma|^{1/2}$ and gets for $f(z)$:
\be
\label{EqfzGamNegat}
 4(1-z)z f' -(1+6z)f  -  p\, z^{1/2} + 1=0
\ee
The solution is generally singular at $z=1$ and the branch for $0<z<1$ (which includes the asymptotic region since $r^4 \sim 1/z$) is obtained similarly to the case $\gamma>0$ above to be:
\be
\label{SolNegGammafz}
f(z)=\frac{1}{ (1-z)^{7/4}} \left[-\mu z^{1/4}+p z^{1/2}F\left(-\frac{3}{4},\frac{1}{4},\frac{5}{4},z\right) +F\left(-\frac{3}{4},-\frac{1}{4},\frac{3}{4},z\right) \right]
\ee
and $f(x)$ (now $x=r/(|\gamma|^{1/4}\sqrt{\kappa P})= z^{-1/4}$) is:
 \bea
\label{SolNegGammafx}
f(x)=\frac{1}{ (1-1/x^4)^{7/4}} \left[-\frac{\mu }{x}+\frac{p}{x^2} F\left(-\frac{3}{4},\frac{1}{4},\frac{5}{4},\frac{1}{x^4}\right) +F\left(-\frac{3}{4},-\frac{1}{4},\frac{3}{4},\frac{1}{x^4}\right) \right] \;\; , \;\; x>1
\eea

The asymptotic behavior is still given by Eq. (\ref{Asymptf}) with $\gamma \mapsto -|\gamma|$ and the same goes for the explicit form of $f(r)$ which is still given by Eq. (\ref{Solfr}). However, there is an important difference  with respect to the $\gamma>0$ solutions, which is the singularity of Eq. (\ref{EqfzGamNegat}) and consequently of the generic solutions at $z_s =x_s = 1$ or $r_s = |\gamma|^{1/4}\sqrt{\kappa P}$ . This singularity has a significant effect on the solutions since it is not a coordinate singularity, but a ``real'' curvature singularity at which the Ricci scalar, and the two quadratic invariants $R_{\mu\nu}R^{\mu\nu}$ and $R_{\kappa\lambda\mu\nu}R^{\kappa\lambda\mu\nu}$ all diverge. This means that the $\gamma<0$ BH solutions are defined outside a spherical region of a circumferential radius of $r_s$.

\subsubsection{Adding Cosmolgical Constant}  \label{LambdaTerm}
The cosmological constant modifies the field equations in a very simple way such that only Eq. (\ref{Eqf}) gets modified so the two basic Eqs. (\ref{Eqa})--(\ref{Eqf}) become:
\be
\label{EqaCosConst}
 (r^4+\gamma\kappa^2 P^2)r\frac{a'}{a}  + 3\gamma\kappa^2 P^2 = 0
\ee
\be
\label{EqfCosConst}
 (r^4+\gamma\kappa^2 P^2)rf'   +(r^4-6\gamma\kappa^2 P^2)f+  \frac{\kappa P^2}{2}r^2-r^4 +  \frac{\Lambda}{2}r^6=0
\ee
Thus, only $f(r)$ is modified and the solution is obtained along the same lines as before to be:
  \bea
\label{SolfrLam} \nonumber
f(r)=\left(1+ \frac{\gamma\kappa^2 P^2}{r^4}\right)^{-7/4} \left[ F\left(-\frac{3}{4},-\frac{1}{4},\frac{3}{4},-\frac{\gamma\kappa^2 P^2}{r^4}\right)-\frac{2M}{r}+\frac{\kappa P^2}{2 r^2} F\left(-\frac{3}{4},\frac{1}{4},\frac{5}{4},-\frac{\gamma\kappa^2 P^2}{r^4}\right) -  \right. \nonumber \\ \left.
\frac{\Lambda r^2}{3} F\left(-\frac{3}{4},-\frac{3}{4},\frac{1}{4},-\frac{\gamma\kappa^2 P^2}{r^4}\right) \right]\hspace{0.3cm}
\eea
The last ``cosmological'' term of $f(r)$ has generally the ordinary behavior as for $\gamma=0$: asymptotically it increases or decreases as $\pm r^2$ according to the sign of $\Lambda$ (or actually of $-\Lambda$). Also, it vanishes at the origin as when the Horndeski term is absent. Still there are differences, since unlike the $\gamma=0$ case it does not have a definite sign for all $r$, so it may play a role in the horizon structure, i.e. the zeroes of $f(r)$. Studying this system for non-vanishing cosmological constant is outside the scope of this work. From now on we will focus on the asymptotically-flat solutions and defer the study of the effect of non-zero cosmological constant to a future publication.


\begin{figure}[t!!]
\begin{center}
{\includegraphics[width=8cm, angle = -00]{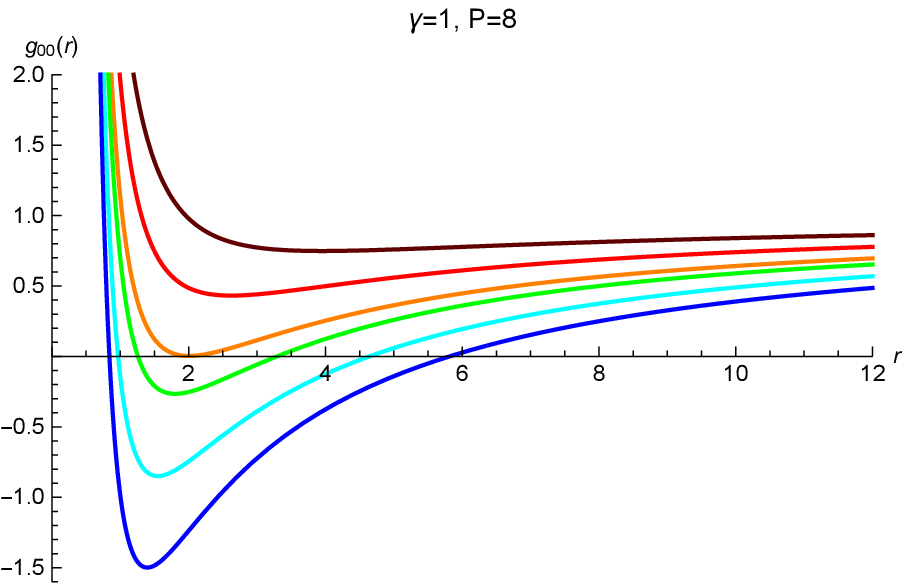}}
{\includegraphics[width=8cm, angle = -00]{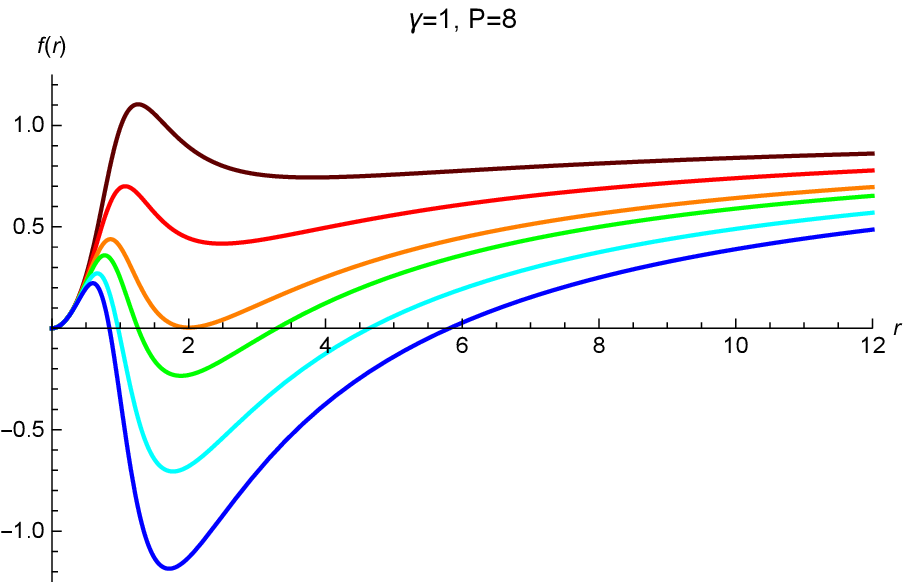}}
{\includegraphics[width=8cm, angle = -00]{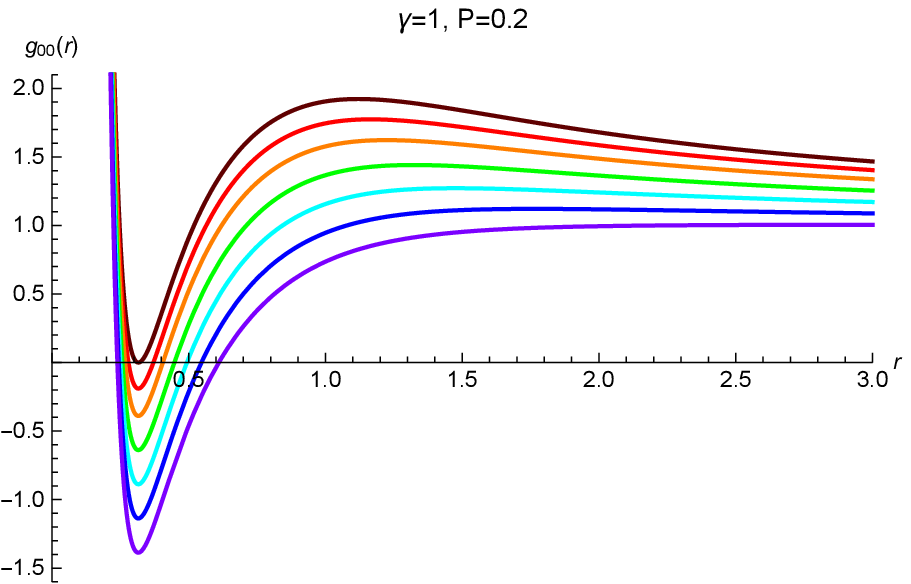}}
{\includegraphics[width=8cm, angle = -00]{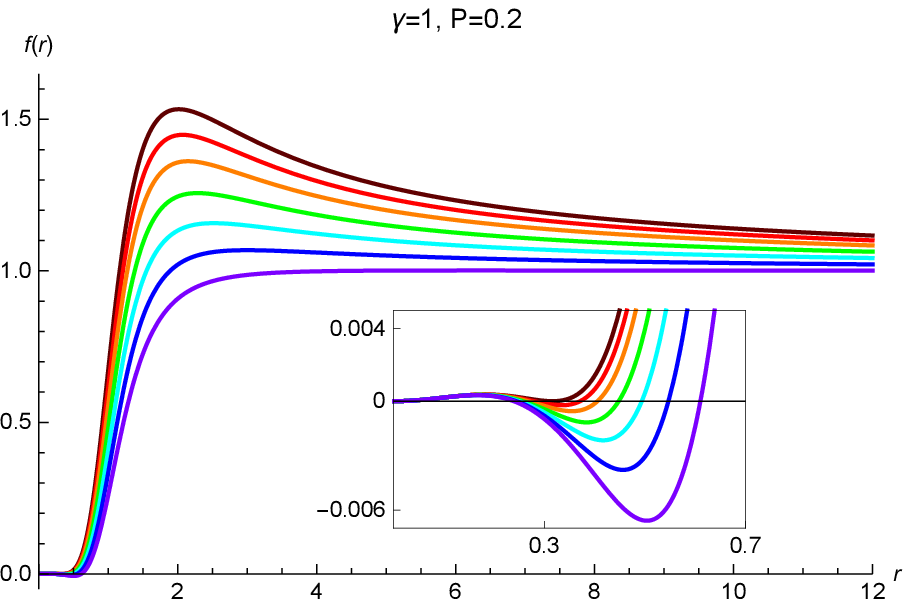}}
\end{center}
\caption{{\small Upper part: Profiles of  $g_{00}(r)$ and of $f(r)=-1/g_{rr}(r)$ for $\gamma=1$ and $P=8$ for several values of the dimensionless mass parameter: $2M/r_s =\mu=2,\;3,\; 3.988 \;  (for\; the\; extremal\; solution),\; 4.5,\; 5.5,\; 6.5$.  The mass increases in a ``spectral order'' from red to blue, or lower curves correspond to larger mass. The two smaller mass naked singularity solutions are unphysical}. Lower part: Profiles of negative mass BHs which appear for small $P$ (see text). The $\mu$ values are: $\mu=-1.390$ (\textit{extremal}), $-1.2,\;-1.0,\;-0.75,\;-0.5,\;-0.25,\;0$. Note especially the decreasing $g_{00}(r)$ beyond the maximum. The radial coordinate is rescaled as $r/r_s$.} \label{FigProfiles}
\end{figure}

\section{Main Characteristics of the Solutions: $\gamma>0$}\label{CharactreisticsSolGamPos}
\setcounter{equation}{0}

\begin{figure}[t!!]
\begin{center}
{\includegraphics[width=8.5cm, angle = -00]{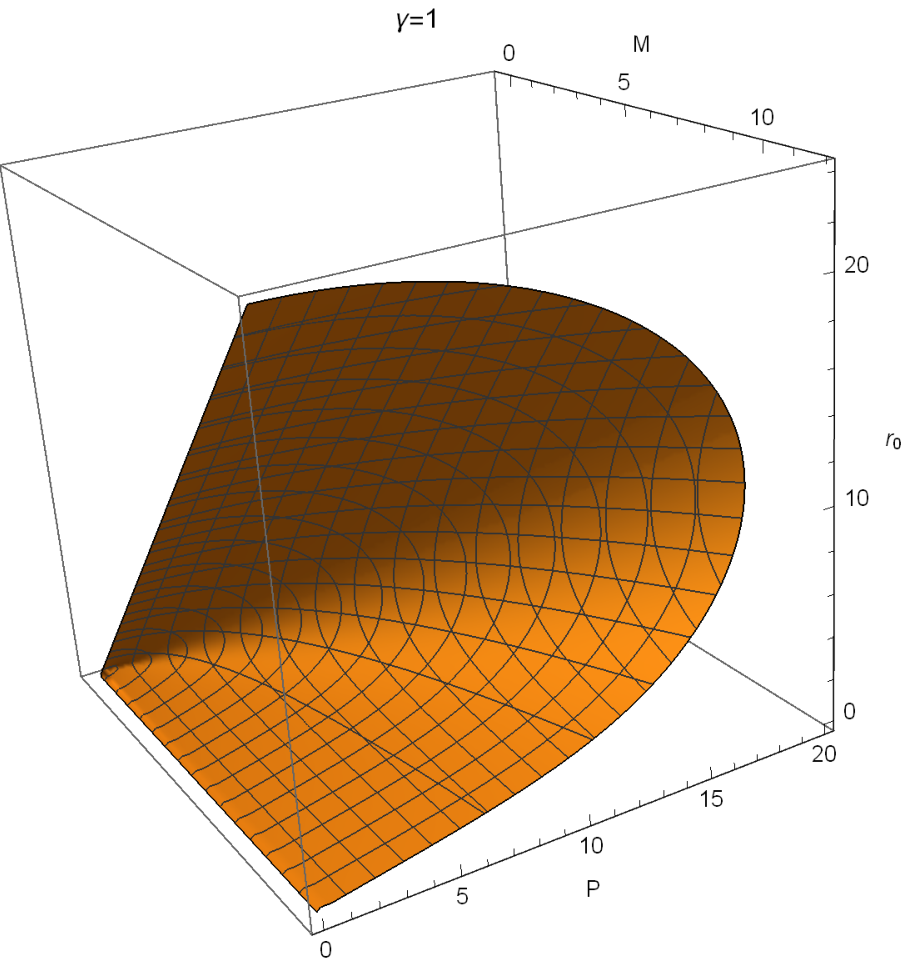}}
{\includegraphics[width=7.5cm, angle = -00]{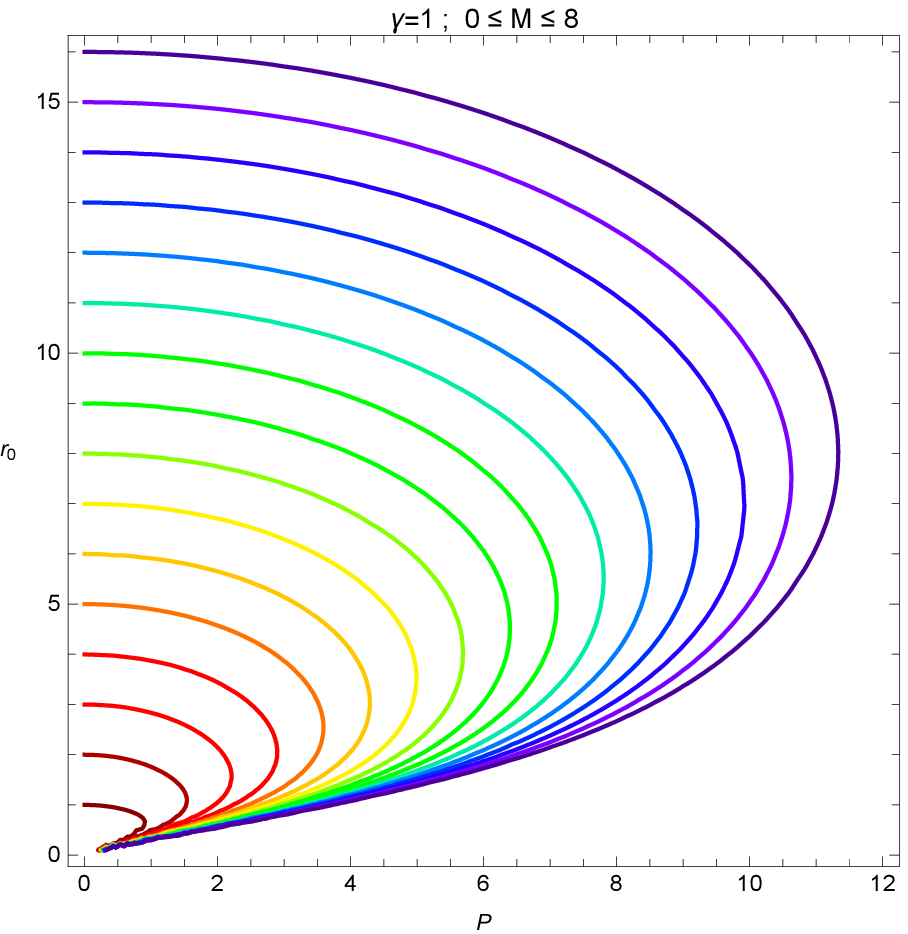}}
\end{center}
\caption{{\small Left: Dependence of $r_0$ on the mass and magnetic charge for $\gamma=1$. Right: Sections of the surface at several values of $M$ in even intervals from $M=0$ to $M=8$. The mass values increase in a ``spectral order'' from red to violet. The radial coordinate and the mass here are rescaled by $\sqrt{\kappa}$.}}\label{FigrHvsMP2+3D}
\end{figure}

We start by inspection of the general structure of the metric components of the solutions. Figure \ref{FigProfiles}  shows profiles of the metric components for the typical value  $\gamma=1$ of the coupling constant. The upper two panels of the figure show the common pattern that is found throughout most of parameter space where $\gamma>0$. In these profile plots we rescale the radial coordinate by the length parameter $r_s$, so we actually use the dimensionless coordinate $x=r/r_s$. In the next plots we will rescale the radial coordinate, mass and temperature by $\sqrt{\kappa}$ only in order to identify more clearly the role of each of the mass, magnetic charge and horizon separately. Unless ambiguity may occur, we will use the same symbols for the unrescaled as well as rescaled quantities of both kinds.

First we note that the function $f(r)$ vanishes at the origin, unlike the RN case. This is a direct result from the modification of the RN solution by the hypergeometric factors. It is obvious that away from the origin and asymptotically, the behavior is similar to RN, but near the origin the behavior is modified drastically. A straightforward calculation yields the following expansion near the origin: $f(r) = r^2/8\gamma\kappa - r^4/2\gamma (\kappa P_0)^2 + \cdots$ . In order to understand better the behavior near the origin we expand also the mass function $M(r)$ and find  $M(r) = r/2 - r^3/(16\gamma \kappa)  + \cdots$ which implies $M(0)=0$. So it seems that these BHs do not have a point mass at $r=0$. Still the origin is a point of  curvature singularity at which the Ricci scalar, and the two quadratic invariants $R_{\mu\nu}R^{\mu\nu}$ and $R_{\kappa\lambda\mu\nu}R^{\kappa\lambda\mu\nu}$ all diverge. The reason for this singularity is the diverging mass density which behaves near the origin like $T_0^0  = 1/\kappa r^2 + \cdots$ .

Second, we turn to study the horizon structure of the solutions.  In addition to the zero at the origin, $f(r)$ may have two  nodes for $r>0$ or no nodes, and in between, there is the special (``extremal'') case where the two nodes degenerate to one. The metric component $g_{00}(r)$ has similar behavior to the corresponding RN metric, so generally the horizon structure is similar to RN. The nodeless solutions violate Cosmic Censorship and will be usually discarded.
Figure \ref{FigrHvsMP2+3D} summarizes the general dependence of the points where $f(r_0)=0$ on the mass and magnetic charge. Note that the function is doubly-valued. The larger of the two zeroes (when they exist) is of course the event  horizon of the black hole which we will denote by $r_H$. Otherwise, we have naked singularities. The special value of $r_0$ where the two zeroes merge corresponds to the extremal BH. From the fact that for an extremal BH both $f(r)$ and $f'(r)$ vanish at $r_0$, it can be deduced that the extremal BH radius is determined by the magnetic charge as $r_{ext}=\sqrt{\kappa P^2/2}$ as in the RN case. Unlike the RN case, the extremal mass is not equal to $\sqrt{\kappa P^2/2}$. It is smaller as seen e.g. from the extremal mass value cited in Fig. \ref{FigProfiles} which is smaller than $\sqrt{\kappa P^2/2}$.

The simplest method to obtain all those results is to get from the equation $f(r_0 )=0$ an explicit expression for $M(\gamma, P , r_0)$:
\be
M(\gamma, P , r_0)=  \frac{r_0}{2} F\left(-\frac{3}{4},-\frac{1}{4},\frac{3}{4},-\frac{\gamma\kappa^2 P^2}{ r_0^4}\right) +
\frac{\kappa P^2}{ 4r_0} F\left(-\frac{3}{4},\frac{1}{4},\frac{5}{4},-\frac{\gamma\kappa^2 P^2}{ r_0^4}\right)
 \label{BHMass}
\ee
Differentiating $M(\gamma, P , r_0)$ with respect to $r_0$ (using (\ref{HyperGeomIdent})) in order to find the extremal point gives directly the linear relation $r_{ext}=\sqrt{\kappa P^2/2}$ (or $x_{ext}=p^{1/2}$). Substituting this relation back in Eq.(\ref{BHMass}) yield the equation for the extremal mass curve in the $P$-$M$ plane:
\be
M_{ext}(\gamma, P)= \sqrt{\frac{\kappa P^2}{8}}\;\left[ F\left(-\frac{3}{4},-\frac{1}{4},\frac{3}{4},-\frac{4\gamma }{ P^2}\right) +
 F\left(-\frac{3}{4},\frac{1}{4},\frac{5}{4},-\frac{4\gamma }{ P^2}\right)\right]
 \label{ExtremalBHMass}
\ee
It is quite easy to see from this expression that $M_{ext}(\gamma, P)$ is always (for $\gamma>0$) smaller than the  corresponding RN value and that the difference decreases with increasing magnetic charge.

So we find it  illuminating to add the other two sections of the three-dimensional surface of Fig. \ref{FigrHvsMP2+3D} which are presented below in Fig. \ref{Fig2MoreSections}.

\begin{figure}[h!!!]
\begin{center}
{\includegraphics[width=8.0cm, angle = -00]{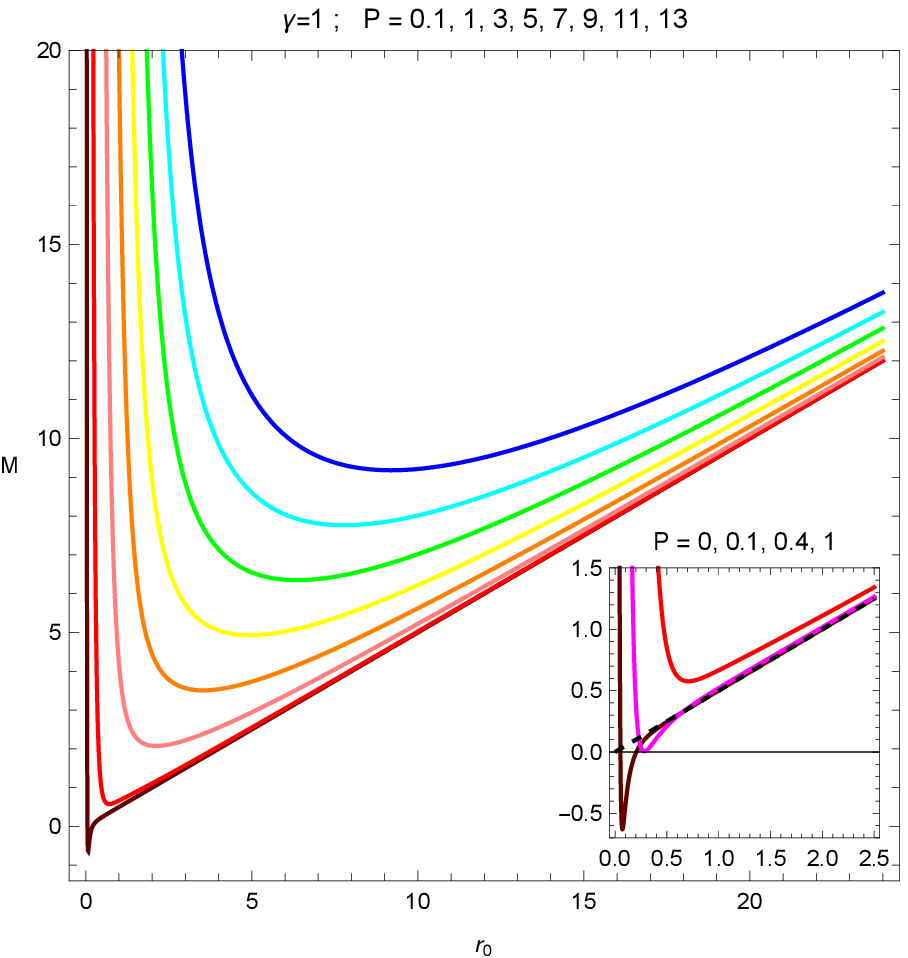}}
{\includegraphics[width=8.0cm, angle = -00]{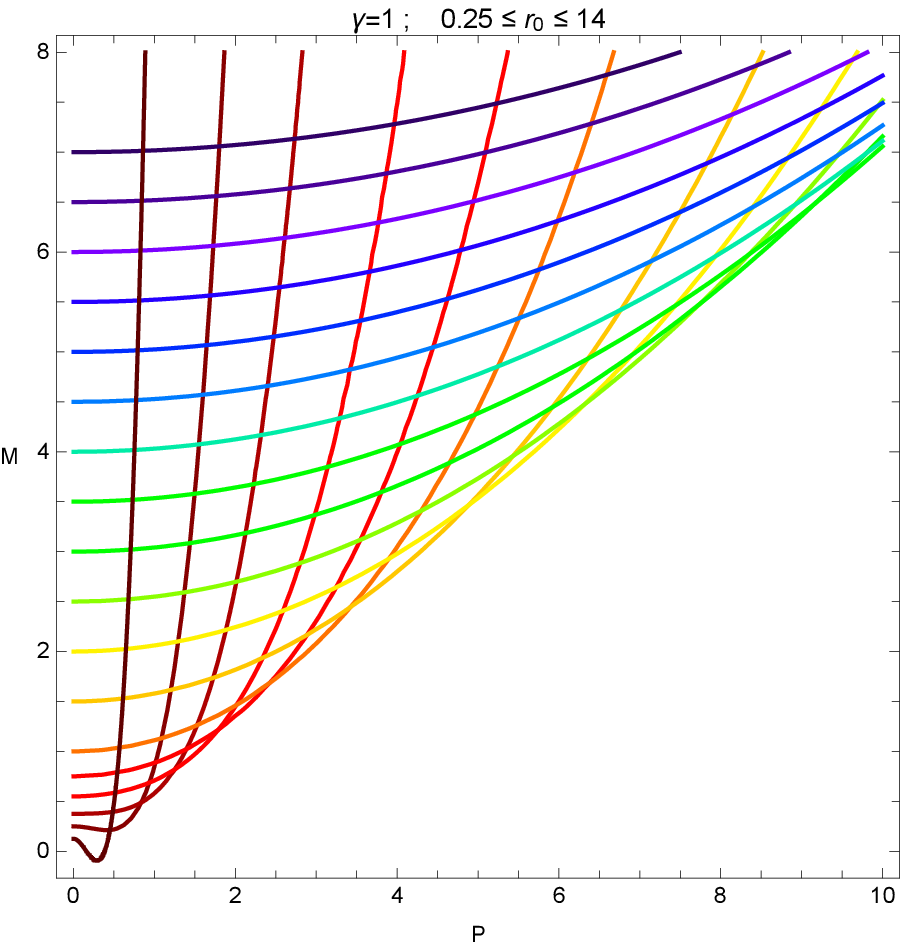}}
\end{center}
\caption{{\small The two other sections of the surface in the $M$-$P$-$r_0$ space of Fig. \ref{FigrHvsMP2+3D}.
Left: $r_0$-$M$ section for several values of $P$. Right: $P$-$M$ section for several values of $r_0$. Note the negative mass region near the origin in both plots. The black dashed curve in the left panel (the insert) corresponds to the \textbf{S} case $P=0$. The radial coordinate and the mass here are rescaled by $\sqrt{\kappa}$.}}\label{Fig2MoreSections}
\end{figure}

A new feature of this Einstein-Maxwell-Horndeski system is the existence of BH-like solutions with negative mass as is obvious from the lower two panels of Fig. \ref{FigProfiles} and from
both parts of Fig. \ref{Fig2MoreSections}. Of course, $M=0$ RN or \textbf{S} solutions exist too, but they have naked singularities. As opposed to RN, in the present circumstances  if the magnetic charge is small enough, the singularity is hidden by the two horizons as seen also at the lower left corner of Fig. \ref{FigrHvsMP2+3D} -- where it is obvious that a whole line in the $P$-$r_0$ plane with $M=0$ exists. Moreover, $M=0$ is not a limiting case, but can be crossed and a new type of solutions appears that have the same horizon structure (inner and outer) but presenting a negative mass. This is clearly reflected in the $g_{00}(r)$ metric component of Fig. \ref{FigProfiles} which asymptotically decreases with $r$. So these solutions have a repulsive gravitational field. This of course has no analogue in the RN solution where the horizon structure is determined by the mass and charge such that $M$ should be not only positive, but also larger than $\sqrt{\kappa P^2/2}$. In the present case there is an additional Horndeski contribution to the extremal mass and it is easy to see that it is negative and its absolute value increases with $\gamma$.

The domain where negative BH masses are possible can be obtained from Eq.  (\ref{ExtremalBHMass}) for the extremal mass, by the condition $M_{ext}<0$. This translates to a transcendental algebraic equation in the sum of the two hypergeometric functions which gives the maximal $\gamma$-dependent  value of $P$ where negative mass BHs are possible, to satisfy $4\gamma / P^2 = 25.61190$. This means that negative mass BHs can be found for magnetic charges of $0 < P < 0.395193 \gamma^{1/2}$. Note that for any  $P$ in this interval the BH masses may be negative, but they are still bounded from below. $P=0$ is a singular limit in the sense that it corresponds to the \textbf{S} case where $M(r_0 )=r_0 /2$.

\begin{figure}[t!!]
\begin{center}
{\includegraphics[width=6.50cm, angle = -00]{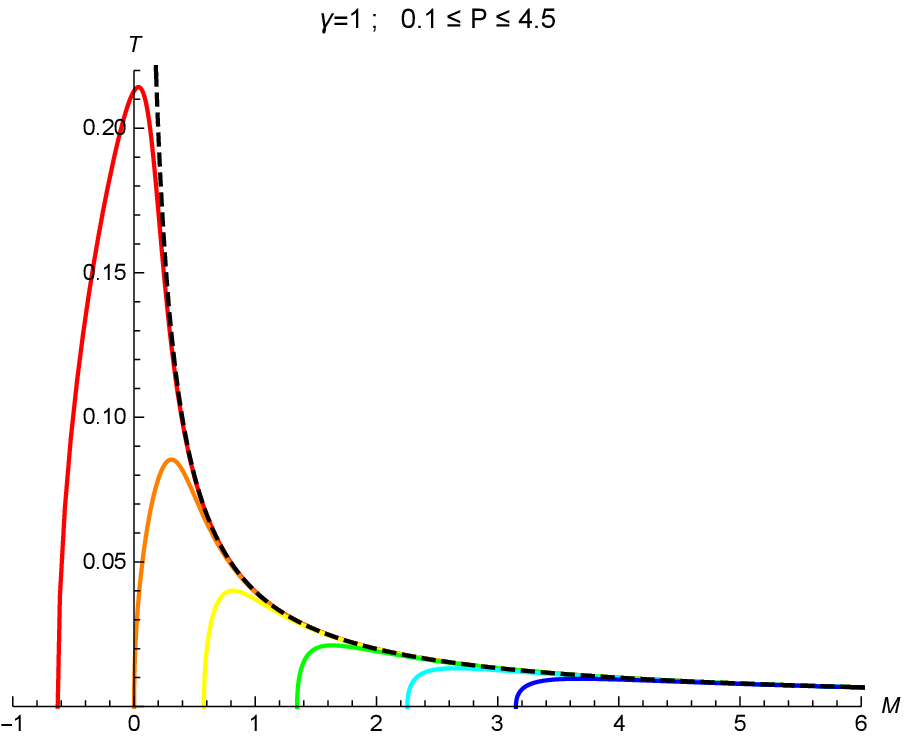}}
{\includegraphics[width=9.50cm, angle = -00]{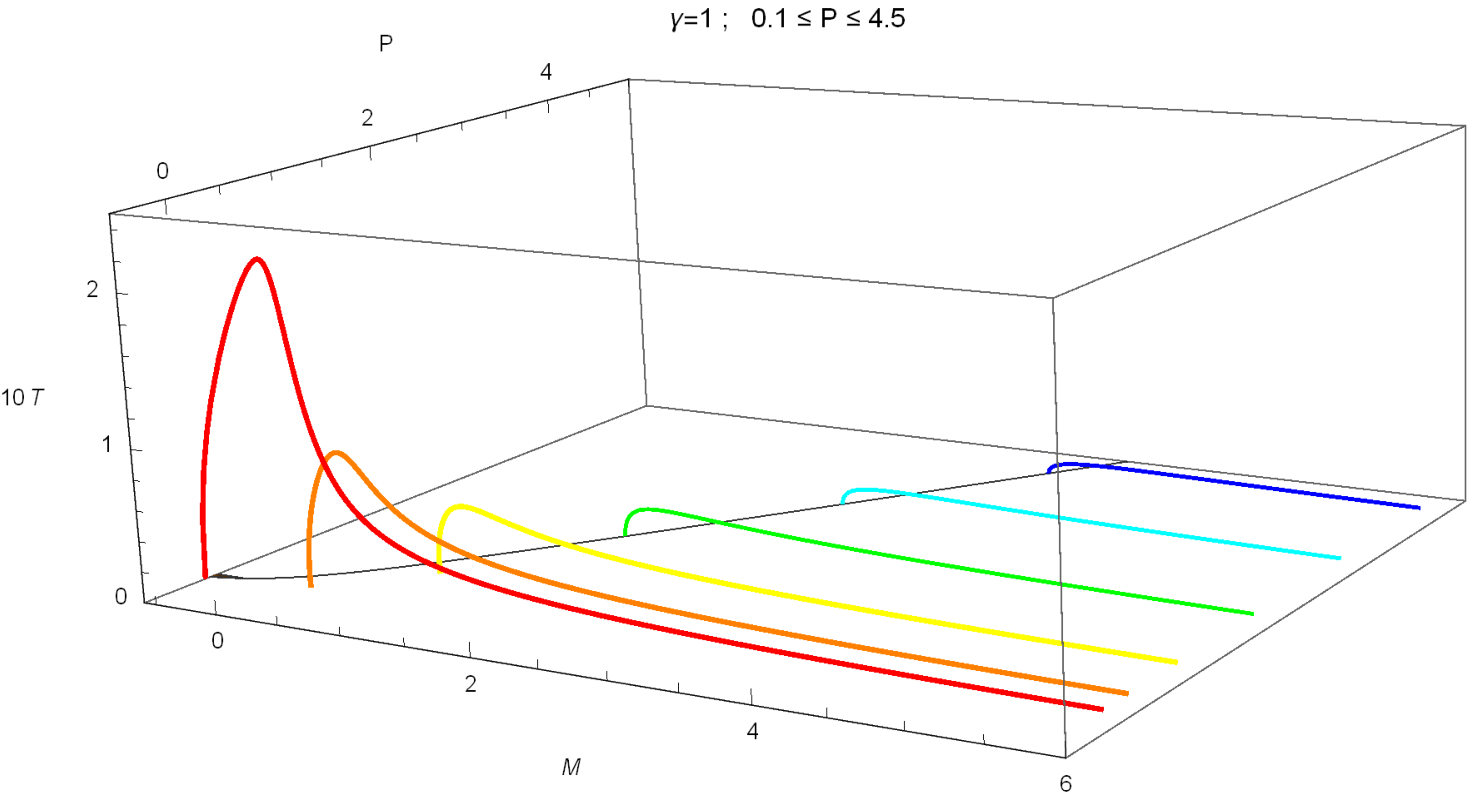}}
\end{center}
\caption{{\small The BH temperature as a function of the BH mass for several values of the magnetic charge. All curves start at the extremal BHs where $T=0$.  The dashed line in the LHS panel corresponds to the \textbf{S} BH. The line in the RHS panel corresponds to Eq. (\ref{ExtremalBHMass}).}}\label{FigTemperature}
\end{figure}

Next, we turn to  the temperature of these black holes, for which we adopt the conventional definition \cite{WaldGR} in terms of the surface gravity ${\cal K}$: $T={\cal K}/2\pi$. In the static spherically-symmetric case with our parametrization, the temperature is given by $T(\gamma, P , r_H)= a(r_H)f'(r_H) /(4\pi)$ and by using the field equations we find the explicit result
\be
T(\gamma, P , r_H)= \frac{r_H^2-\kappa P^2 /2}{4 \pi  r_H^2 (r_H^4+\gamma\kappa^2 P^2)^{1/4}}.
 \label{BHTemperature}
\ee
First we note that the extremal BHs have zero temperature as usual. We also can check that when the BH is not magnetically charged we get back the Bekenstein-Hawking temperature $T=1/4 \pi  r_H$. More generally, we can learn how the BH temperature depends on the mass and magnetic charge from Fig. \ref{FigTemperature}. The function $T(\gamma, P , M)$ cannot be written down explicitly, so the best we can do is to use the parametric representation $(M(\gamma, P , r_H), T(\gamma, P , r_H))$ with $r_H$ as a parameter. The most prominent feature of all curves is the existence of a maximal temperature, very much like the behavior of the RN BH. However, while the maximal RN temperature  is inversely proportional to the magnetic charge, in the presence of the Horndeski term it is not exactly $\sim 1/P$ for all $P$, although the difference is quite small and the behavior becomes $ 1/P$ asymptotically.

This brings us to the final point which is the effect of varying the non-minimal coupling parameter $\gamma$. It has of course a decisive effect, but technically it is quite simple to understand since most of it is realized through a scaling behavior which originates from the fact that there are actually two independent free parameters which determine the solutions: $\mu$ and $p$. So the effect of $\gamma$ is done only through these two parameters. One example of the scaling behavior can be seen in Eq. (\ref{ExtremalBHMass}) for the extremal mass where $M_{ext}(\gamma, P)$ depends on $\gamma$ through the ratio $\gamma / P^2$. This scaling behavior is not valid for  vanishing magnetic charge, but we may exclude this case in the present discussion since anyhow we know already that $P=0$ gives the Schwarzschild (\textbf{S}) solution for vanishing as well as for non-vanishing $\gamma$.

The same can be done for the temperature if we use the radial coordinate $\xi=r/r_{ext}=r\sqrt{2/\kappa P^2}$. Then we can write the temperature as

\be
T(\gamma, P , \xi_H)=\frac{\sqrt{2}}{4 \pi \sqrt{\kappa P^2}} \frac{\xi_H^2-1}{ \xi_H^2 (\xi_H^4+4\gamma/ P^2)^{1/4}}.
 \label{BHTemperatureDimlss}
\ee
So here too, most of the effect of increasing $\gamma$ can be done also by decreasing the magnetic field and keeping $\gamma$ fixed -- of course, as long as it stays finite.

\begin{figure}[b!!]
\begin{center}
{\includegraphics[width=8cm, angle = -00]{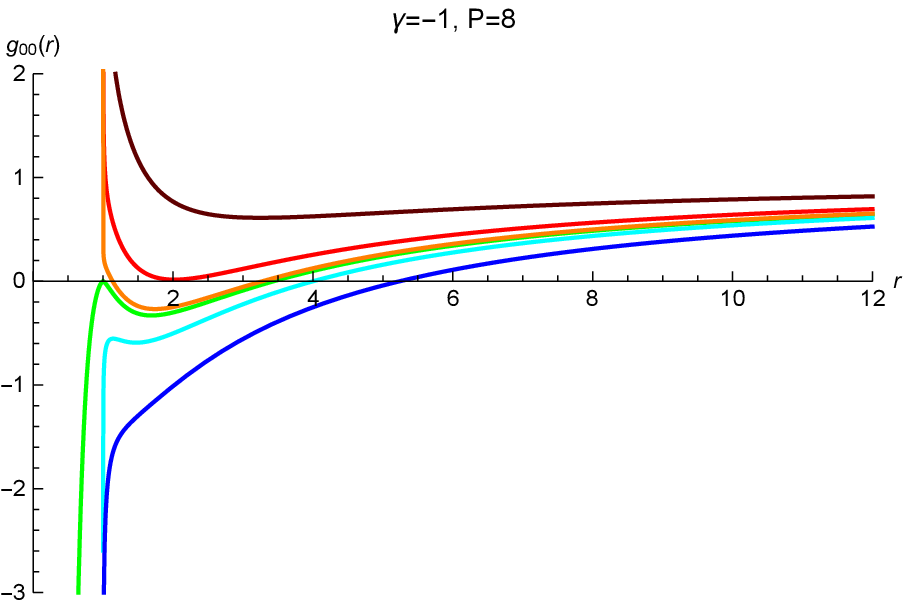}}
{\includegraphics[width=8cm, angle = -00]{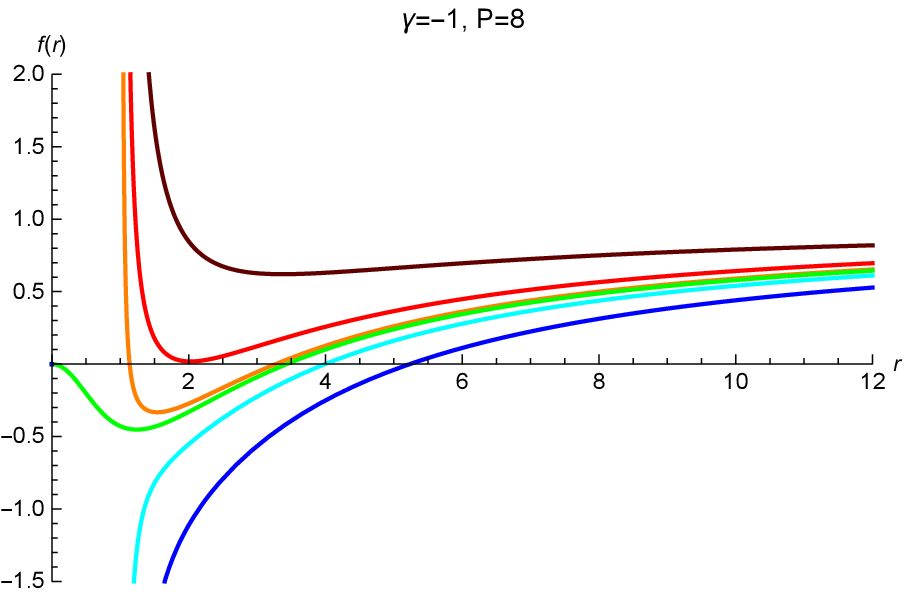}}
\end{center}
\caption{{\small Profiles of  $g_{00}(r)$ and of $f(r)=-1/g_{rr}(r)$ for $\gamma=-1$ and $P=8$
and several values of the  mass parameter: $\mu=2.5,\;4.013$  (\textit{the extremal solution}), $ 4.5,\; 4.603$   (\textit{the intermediate solution}), $5,\; 6$.}} \label{FigProfilesNegGam}
\end{figure}

\section{Main Characteristics of the Solutions: $\gamma<0$}\label{CharactreisticsSolGamNeg}
\setcounter{equation}{0}
The nature of the solutions for $\gamma<0$ is quite different from that of $\gamma>0$, although some similarities do exist. One similarity is the fact that the mass function is still given by Eq. (\ref{BHMass}) and $r_{ext}=\sqrt{\kappa P^2/2}$ still plays the role of the horizon of the Extremal BHs, although the opposite sign changes the behavior noticeably. As for other differences, first we mention again the fact that the $\gamma<0$ solutions are well-defined only outside the singular sphere at $r=r_s$. A direct consequence from this condition is that the horizon size must always be larger than $r_s$, so it seems that  $\gamma<0$ RN-like BH solutions (see below) are more constrained due to the additional condition $r_{ext} \geq  r_s$ which in terms of $P$ reads $ P/2|\gamma|^{1/2}  = p  \geq 1$.

Next, we present in Fig. \ref{FigProfilesNegGam} typical profiles for the same $P$ and $|\gamma|$ as in  Fig. \ref{FigProfiles} and notice immediately that whereas the $\gamma>0$ BHs all have a RN-like behavior (two horizons and $g_{00}(r)\rightarrow \infty$ as $r$ decreases), the $\gamma<0$ ones are split between RN-like behavior for small masses and S-like behavior (single horizon and $g_{00}(r)\rightarrow -\infty$ as $r$ decreases) for large masses. Moreover, $g_{00}(r)$ diverges on the singular sphere $r=r_s$. In between there exists a solution which seems totally finite and regular for all $r>0$. However, this solution too suffers from the same curvature singularity at $r=r_s $ as the calculation of the curvature invariants shows immediately. The ``culprit'' is of course the function $a^{2}(r)$ which for  $\gamma<0$ is differentiable only once at $r=r_s$. We also note that the extremal solution has its double zero at the same point as for $\gamma>0$ but the  mass parameter is higher. Indeed in these two figures the mass parameter is close to $\mu=4$, but it is a result of the large $P$ behavior of the function $M_{ext}(\gamma, P)$ in Eq. (\ref{ExtremalBHMass}). The general pattern is that for $\gamma<0$ the extremal masses are also larger than the RN bound of $\sqrt{\kappa P^2/2}$.  This is in accord with the previous observation that increasing $\gamma$ tends to decrease the BH mass.

\begin{figure}[b!!]
\begin{center}
{\includegraphics[width=8.00cm, angle = -00]{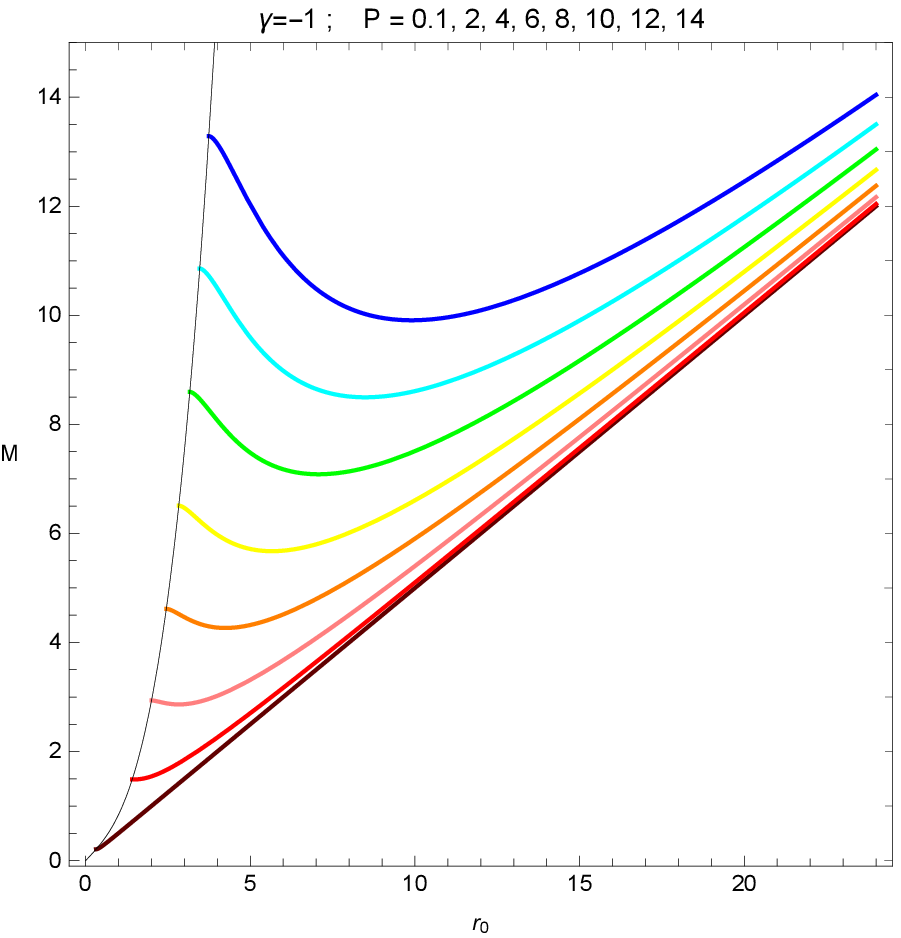}}
{\includegraphics[width=8.00cm, angle = -00]{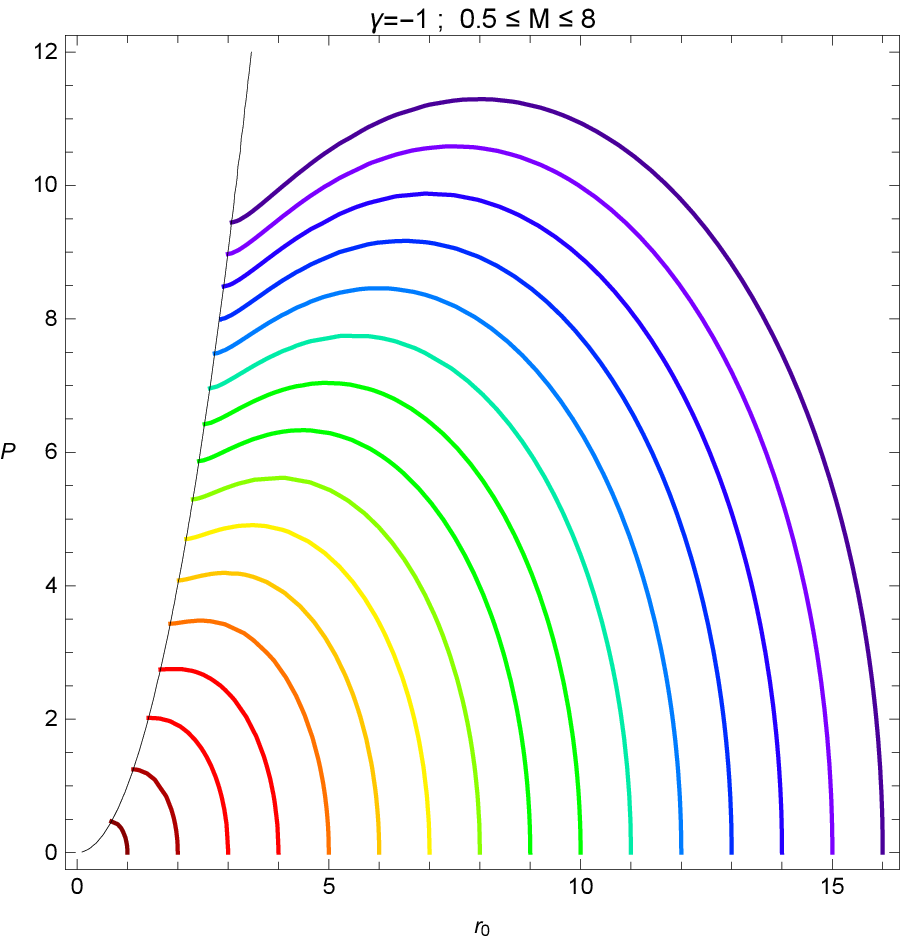}}
\end{center}
\caption{{\small BH mass and magnetic charge vs. $r_0$ for $\gamma = -1$ and several values of $P$ and $M$ respectively. The line at the left in each panel represents the curvature singularity which is located at $x=r/(\gamma^{1/4}\sqrt{\kappa P}) = 1$.}}\label{FigMvsrH-NegGam}
\end{figure}

The same function of (\ref{ExtremalBHMass}) can give an additional view on the fact that the RN-like solutions exist in a limited domain in parameter space: First of all we recall that $M_{ext}(\gamma , P)$ is defined for RN-like solutions only. It says nothing about S-like ones. Second, $M_{ext}(\gamma , P)$ is defined (actually, is real) only for $-4\gamma/P^2 < 1$ which for $\gamma<0$ is the same as $ p  > 1$. So extremal RN BHs can exist for $p>1$ only. The condition $p<1$ is realized by the S-like BHs. In other words, if $p<1$, all BH solutions are S-like. This points up to another special feature of these S-like BHs: since the horizon size cannot be arbitrarily small, there is a lower bound on the mass of the S-like BHs with a given $P$ provided also that $P<2|\gamma|^{1/2}$:

\be
M^{(S)}_{min}(\gamma, P)=   \Gamma \left(\frac{7}{4}\right) |\gamma|^{1/4} \sqrt{\kappa P} \left( \frac{\Gamma(3/4)}{\sqrt{\pi}} + \frac{\Gamma(5/4) P}{4|\gamma|^{1/2}} \right) \;\;\;\;  , \;\; P<2|\gamma|^{1/2}
 \label{S-like-Mass-Minimum}
\ee
This is the analogue of the S-like BHs to the extremal mass given by Eq. (\ref{ExtremalBHMass}).

To return to $p>1$, this is not a sufficient condition to have RN-like solutions. The reason is more obvious from Fig. \ref{FigMvsrH-NegGam}  which presents the dependence of the BH mass and magnetic charge on the zeroes $r_0$ of the metric components. Notice the ``forbidden zone'' bounded by the curvature singularity. Inspecting the left-hand-side panel of Fig. \ref{FigMvsrH-NegGam} we can see that above the $p=1$ line, the curves of $M(\gamma, P, r_0)$ develop a minimum (at a certain $r_0$), but unlike the RN case where for each $M$ there are two values of $r_0$, in the present case there is an additional upper bound, say $M_{inter}(\gamma, P)$ above which all BH solutions will have a single zero. This $M_{inter}(\gamma, P)$ is the mass where the curve $M(\gamma, P, r_0)$ crosses the line of curvature singularity. For the case depicted in  Fig. \ref{FigProfilesNegGam} with $p=4$, the corresponding mass parameter is $\mu_{inter} = 4.603$ which determines the intermediate solution shown in the figure. So, RN-like solutions exist only for masses $M_{ext}(\gamma, P)<M<M_{inter}(\gamma, P)$. If $M > M_{inter}(\gamma, P)$ there exist only S-like BHs, as is seen clearly in Fig. \ref{FigProfilesNegGam}.

Notice also in Fig. \ref{FigMvsrH-NegGam} the region of small $M$ and $P$ where only one zero exists for each $M$ and $P$. This is the region of $p<1$ where all BHs are S-like. We did not include a figure containing profiles corresponding to this case. They are similar to those of the lower curves of Fig. \ref{FigProfilesNegGam}.

Also obvious from these plots (and from direct study of (\ref{BHMass}) for $\gamma<0$) is that no negative mass BHs exist for  $\gamma<0$.


Next we move to the temperature which is obtained similarly to (\ref{BHTemperature})  as

\be
T(\gamma<0, P , r_H)= \frac{r_H^2-\kappa P^2 /2}{4 \pi  r_H^2 (r_H^4-|\gamma|\kappa^2 P^2)^{1/4}}.
 \label{BHTemperatureNegGamLargeH}
\ee
Of course, in order to stay away from the singularity at $r=r_s$ we have to assume $r_H>r_s$

\begin{figure}[h!!!]
\begin{center}
{\includegraphics[width=8.00cm, angle = -00]{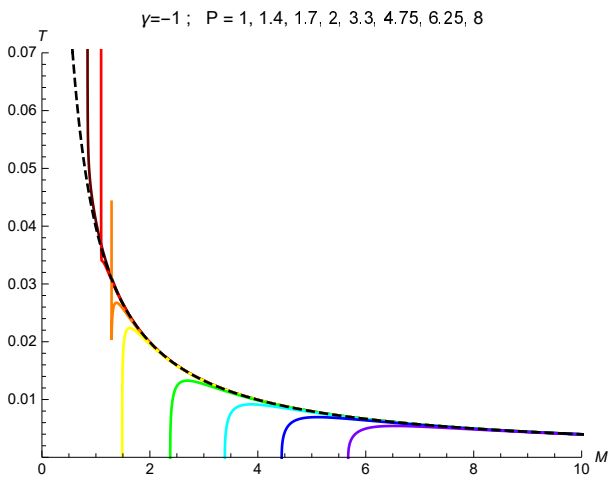}}
\end{center}
\caption{{\small The BH temperature as a function of the BH mass for several values of the magnetic charge. Notice the two different types of behavior: The RN-like curves start at the extremal BHs where $T=0$, attain a maximum and decrease. The others correspond to the small $P$ S-like BHs. The ``spikes'' are not singularities, but look like that because of resolution.  The dashed line corresponds to the S BH.}}\label{FigTemperatureNegGamma}
\end{figure}

 We expect to find for the temperature two different types of behavior corresponding to the two types of $\gamma < 0$ BH solutions, and indeed this is the case as demonstrated by Fig. \ref{FigTemperatureNegGamma} which presents the mass dependence of the temperature for several values of $P$. For large enough magnetic charge the temperature  behaves similarly to the $\gamma>0$ RN-like case:   it starts with $T(r_{ext})=0$ rises to a maximum and then decreases monotonically, asymptotically to zero. For the smaller values of $P$, no extremal BHs exist since the BH solutions are S-like. Thus the temperature is always strictly positive and generally decreases with $M$. The decrease is monotonic for small enough $P$, while some ``spikes'' develop for higher values of $P$ which are still too small to allow for extremal BHs, i.e. still $p<1$. These ``spikes'' are not singularities, but very narrow minima. However, the temperature diverges as $r_H \downarrow r_s$, or rather as  $M \downarrow M(\gamma<0, P , r_s)$.

 By comparison with Fig. \ref{FigTemperature} we find that for the same values of mass and charge, the  temperatures of the $\gamma<0$ RN-like BHs are very close to the $\gamma>0$ ones, but not identical.

\section{Geodesics of Neutral Particles}  \label{GeodesicsNeut}
\setcounter{equation}{0}
The trajectories $x^\mu (\tau)$ of test particles around the magnetic BH are determined by the Lagrangian
\be
L=-m\sqrt{a^2 f(r)\dot{t}^2-\dot{r}^2/f(r)-r^2(\dot{\theta}^2+\sin^2\theta~\dot{\phi}^2)}-qP(1-\cos\theta)\dot{\phi}
\label{GeodEqsMagLagrangian}\ee
where $\tau$ is proper time and $q$ the test charge.

First we concentrate on the geodesic equations for neutral particles ($q=0$; either time-like or null). In a static spherically-symmetric spacetime the motion is planar and without loss of generality the orbital plane can be taken as $\theta = \pi /2$. The equations of motion may be reduced to the two following first order equations which we write for the dimensionless radial coordinate $x(\tau)=r(\tau)/r_s$ and the azimuthal angle $\phi(\tau)$ as:
\be
x^2 \dot{\phi} = \ell \,\,\, ; \,\,\,\,\,\, \dot{x}^2+f(x)\left( \epsilon +\frac{\ell^2}{x^2} \right) - \frac{{\cal E}^2}{a^2(x)} = 0
\label{GeodEqsMagIntegrated}\ee
 The parameter $\epsilon$ takes the value 1 for timelike geodesics of massive particles, but may cover also null geodesics of massless particles if $\epsilon = 0$ and an affine parameter is used instead of proper time. The parameters $\ell$ and ${\cal E}$ are the rescaled conserved angular momentum and energy of the particle. The two terms in the radial equation act as an effective potential (we omit the factor of 2 from classical mechanics) $V_{eff} (x)$ which is very useful to classify the possible orbits and trajectories:
 \be
V_{eff} (x)=f(x)\left( \epsilon +\frac{\ell^2}{x^2} \right) - \frac{{\cal E}^2}{a^2(x)} .
\label{EffPotNeut} \ee
First we notice that the energy (squared) does not appear here as an additive constant, but as a parameter in the effective potential. The ``effective energy'' of the trajectories is always zero. Still ${\cal E}^2$ controls the large distance behavior of the trajectories similarly to the more well-known cases of \textbf{S} and RN where $a(x)=1$, since asymptotically $V_{eff}(x)\rightarrow  \epsilon -{\cal E}^2$. Thus, for ${\cal E}^2\geq\epsilon$ there will be open  trajectories, i.e., they can reach $x\rightarrow\infty$. If ${\cal E}^2<\epsilon$, the motion is bounded by a maximal radial distance, since the radial velocity vanishes at some finite value of $x$. This last condition cannot be realized for light ($\epsilon=0$), so light rays are open, except in the special case (if it exists) of circular motion ($\dot{x}(\tau)=0$). These general two types do not exhaust all possibilities. For that, one has to analyze the effective potential in more detail.
\begin{figure}[b!!]
\begin{center}
{\includegraphics[width=8cm, angle = -00]{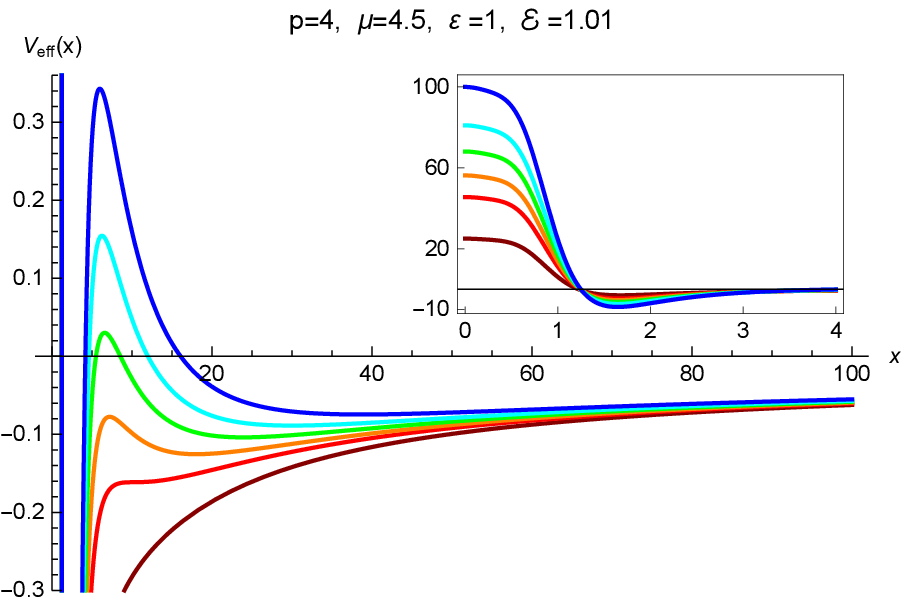}}
{\includegraphics[width=8cm, angle = -00]{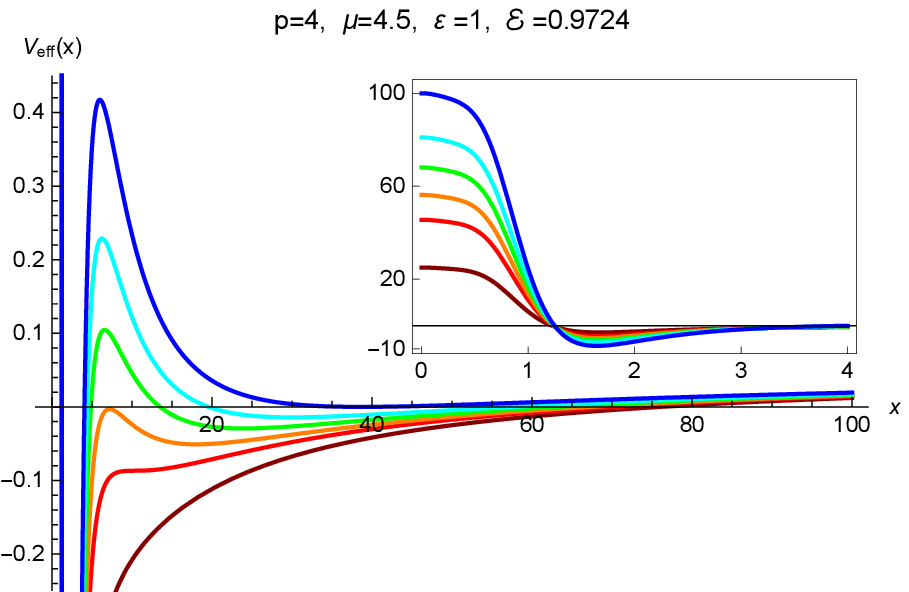}}
\end{center}
\caption{{\small The effective potential for particle trajectories around one of the $\gamma>0$ magnetic HBH solutions presented in Fig. \ref{FigProfiles} for ${\cal E}^2>1$ and ${\cal E}^2<1$ which correspond to unbounded or bounded trajectories respectively. The rescaled angular momentum are:  $\ell = 5,\; 6.75,\; 7.5,\; 8.25,\; 9,\; 10$. The inserts show a ``zoom-out'' of the vicinity of the origin where $V_{eff} (x)$ tends to a constant as $x\rightarrow 0$ unlike the RN case where it diverges. Notice that most of this domain lies within the event horizon which for this case is at $x_H = 3.29$.}} \label{VeffTimelike}
\end{figure}

The explicit form of $V_{eff} (x)$ is obtained by substitution of Eq. (\ref{Solfx}) or (\ref{SolNegGammafx}) according to the sign of $\gamma$ and the  rescaled version of (\ref{Sola}) into (\ref{EffPotNeut}). We will start with $\gamma>0$ and return to the $\gamma<0$ trajectories at the end of this section. An asymptotic expansion of $V_{eff} (x)$ supports the large distance behavior which was described above: $V_{eff} (x) =\epsilon -{\cal E}^2 - \epsilon \mu/x+(\epsilon p+\ell^2)/x^2 +  \cdots$. We also notice that for small values of ${\cal E}$ and $\ell$ the effective potential for particles ($\epsilon = 1$) is generally similar in shape to that of $f(x)$, except near the origin where there is similarity only for vanishing $\ell$. This may be more evident from the expansion for small $x$: $V_{eff} (x) = p\ell^2/4+(\epsilon p-2\ell^2)x^2/4+  \cdots $. From this we also see that $V_{eff}(0)$ is a local maximum for null geodesics and for particles with $\ell^2>p/2$, and a local minimum if $\ell^2<p/2$.

The classification of all the possible motions is straightforward by inspection of the curves of $V_{eff}$ and their dependence on $\ell$ and ${\cal E}$. Fig. \ref{VeffTimelike} demonstrates the main features of the effective potential by showing $V_{eff} (x)$ for massive particles ($\epsilon = 1$) with some representative choices of the parameters. Generally speaking, the number of extremal points of $V_{eff} (x)$ is determined by $\ell$, while the number of its zeroes is determined by ${\cal E}$. Still, these two factors are not completely decoupled and ${\cal E}$ plays also a role on the shape of $V_{eff} (x)$. As mentioned above, the character of the trajectories is determined mainly by the asymptotic value of $V_{eff} (x)$ which is $1-{\cal E}^2$ and by the height of the centrifugal barrier which appears for sufficiently large $\ell$. If ${\cal E}^2>1$ the trajectories are open, i.e., can reach $x\rightarrow\infty$. However, the open trajectories fall further into two subfamilies: If the centrifugal barrier is high enough (large enough $\ell$), they are ``hyperbolic'', i.e. a test particle comes from $x\rightarrow\infty$ and returns to $x\rightarrow\infty$ after reflection from the centrifugal barrier at the ``periastron'' at $x=x_p$. If the centrifugal barrier is not high enough, the trajectories will be ``infalling'', i.e. the test particle will pass over the centrifugal barrier, and will disappear beyond the horizon. If ${\cal E}^2<1$, the motion is bounded, since the radial velocity vanishes at some finite value of $x$. Again the same two possibilities appear: If the centrifugal barrier is high enough, ordinary ``elliptic'' orbits exist. In the opposite case there will be no ``periastron'' and no bound states.

In order to analyze the detailed shape of $V_{eff} (x)$, we obtain an expression for $V'_{eff} (x)$ which we write in terms of $f(x)$ by use of its differential equations and the explicit form of $a(x)$:
 \be
V'_{eff} (x)=-\frac{\left(\ell^2+\epsilon x^2\right) \left(p-x^2\right)}{x
   \left(x^4+1\right)}-\frac{f(x)}{x^3} \left(\frac{\left(x^4-6\right) \left(\ell^2+\epsilon x^2\right)}{x^4+1}+2 \ell^2\right)-\frac{6 {\cal E}^2 x^5}{\left(x^4+1\right)^{5/2}}.
\label{DerivEffPotNeut} \ee
The location of the zeroes of this expression cannot be obtained by a general simple formula as a function of the 4 parameters $p$, $\mu$, ${\cal E}$ and $\ell$ for both values of $\epsilon$. It may be done indirectly from the implicit relation $V'_{eff} (x_\star)=0$ between these quantities. One direct path is to get ${\cal E}^2$ as a function of $p$, $\mu$, $\ell$ and the location of the extremal point $x_\star$:
 \be
{\cal E}^2=\frac{1}{6} \left(x_\star^4+1\right)^{3/2}
\left(\frac{\epsilon }{x_\star^2}+\frac{\ell^2-\epsilon p
   }{x_\star^4}-\frac{p \ell^2}{x_\star^6}-\frac{f(x_\star)}{x_\star^2}\left(\epsilon +\frac{3 \ell^2}{x_\star^2}-\frac{6 \epsilon }{x_\star^4}-\frac{4 \ell^2}{x_\star^6}\right)\right).
\label{EsquareForExtremeVeff} \ee
We will limit our study to BH solutions, i.e. we exclude the solutions where $f(x)$ does not vanish for  $0<x<\infty$. In terms of the mass and magnetic charge, we exclude the region below the line
\be
\mu_{ext}(p) = p^{1/2}\left[  F\left(-\frac{3}{4},-\frac{1}{4},\frac{3}{4},-\frac{1}{ p^2}\right) +
 F\left(-\frac{3}{4},\frac{1}{4},\frac{5}{4},-\frac{1}{ p^2}\right)   \right]
\label{ExtremalBHMassDmlss} \ee
which is the dimensionless form of Eq. (\ref{ExtremalBHMass}). See also the RHS of Fig. \ref{Fig2MoreSections}.
Therefore, for a given BH characterized by the two parameters $p$ and $\mu$, Eq. (\ref{EsquareForExtremeVeff}) becomes a relation between the three other quantities, $\ell$, ${\cal E}$ and $x_\star$, which in principle can be used to give $x_\star$ for any trajectory determined by fixed values of $\ell$ and ${\cal E}$. Of course, there may be more than one extremal point and more than one kind of extremum. In order to infer their nature we calculate the second derivative at the extremal point, $V''_{eff} (x_\star)$ which we write as a ratio:
\bea
V''_{eff} (x_\star) = \Xi (x_\star)/\Upsilon(x_\star) \;\; ,\hspace{12.0cm}\nonumber\\
\Xi (x_\star)= 6 {\cal E}^2 x^6 \left( \left(3 \left(x_\star^4+6\right) x_\star^4+8\right) f(x_\star)-p \left(3 x_\star^4+2\right) x_\star^2+x_\star^8\right)- \hspace{5.4cm} \label{VeffppB}\nonumber\\
2   \epsilon  \left(x_\star^4+1\right)^{5/2} \left(x_\star^2 f(x_\star) \left(p \left(3 x_\star^4-14\right)+\left(12-5 x_\star^4\right) x_\star^2\right)+3
   \left(x_\star^8+2 x_\star^4+8\right) f^2(x_\star)+2 x_\star^4 \left(p-x_\star^2\right)^2\right),\nonumber\\
\Upsilon (x_\star) = x_\star^2 \left(x_\star^4+1\right)^{7/2} \left(\left(3 x_\star^4-4\right)
   f(x_\star)+x_\star^2 \left(p-x_\star^2\right)\right) \hspace{7.4cm}
\label{Veffpp}
\eea
The line $V''_{eff} (x_\star)=0$ separates between the minima ($V''_{eff} (x_\star)>0$) and maxima ($V''_{eff} (x_\star)<0$) and is described by the following relation at the ${\cal E}$-$x_\star$ plane which is obtained directly from Eq. (\ref{VeffppB}):
\bea
{\cal E}^2(V''_{eff} (x_\star)=0)=\hspace{13cm}\\
\frac{\epsilon  \left(x_\star^4+1\right)^{5/2} \left(2 x_\star^4 \left(p-x_\star^2\right)^2+x_\star^2 \left(p \left(3 x_\star^4-14\right)-x_\star^2 \left(5 x_\star^4-12\right)\right)f(x_\star)+3
   \left(x_\star^8+2 x_\star^4+8\right) f^2(x_\star)\right)}{3 \left(3 \left(x_\star^4+6\right) x_\star^4+8\right) x_\star^6 f(x_\star)+3
   x_\star^8 \left(x_\star^6-p \left(3 x_\star^4+2\right)\right)} \nonumber
\label{LineVeffppEq0}
\eea
\begin{figure}[t!!]
\begin{center}
{\includegraphics[width=8.50cm, angle = -00]{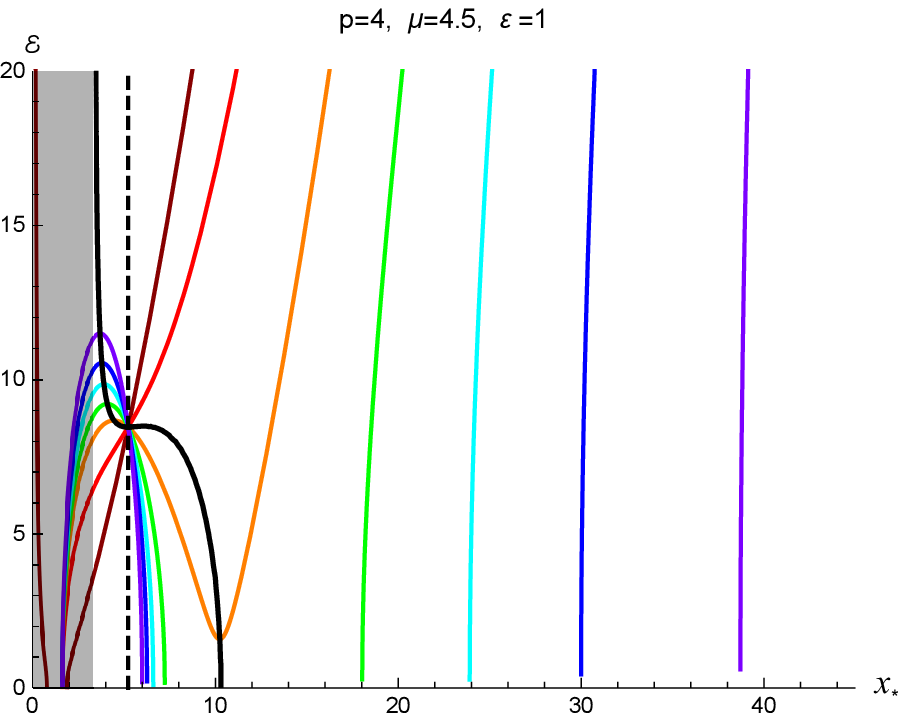}} \hspace{5mm}
{\includegraphics[width=6.50cm, angle = -00]{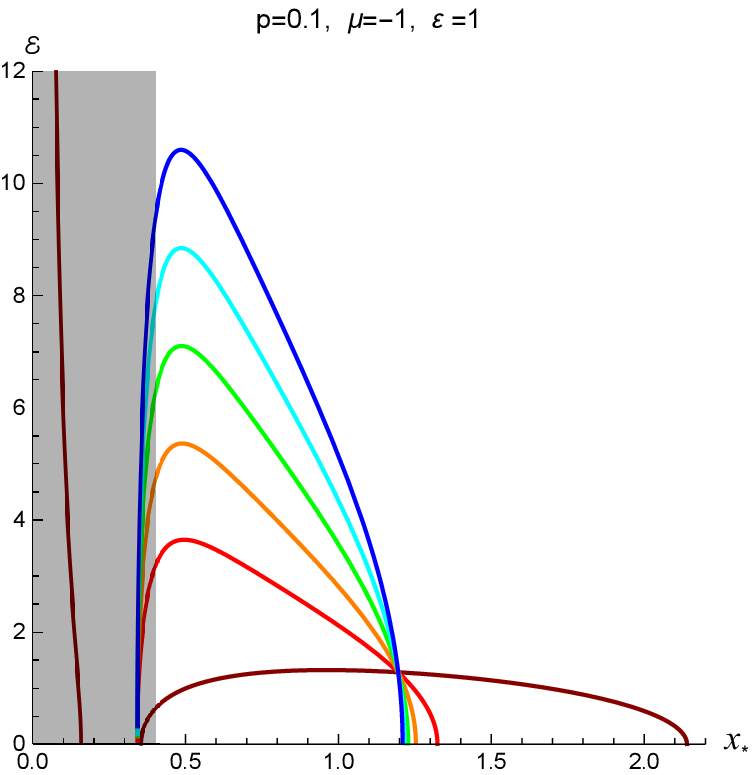}}
\end{center}
\caption{{\small The relation between the extremal locations of $V_{eff} (x)$ and ${\cal E}$ for several values of $\ell$. Left: ordinary MHBHs with the same $\ell$ values as in Fig. \ref{VeffTimelike} (with one addition): $\ell =0,\; 5,\; 6.75,\; 7.5,\; 8.25,\; 9,\; 10$. Right: negative mass MHBH  solutions with $\ell = 0.1,\; 2,\; 3,\; 4,\; 5,\; 6$.  The curves correspond to both maxima and minima of  $V_{eff} (x)$. The maxima and minima alternate, but their nature can be inferred from the fact that the most exterior one (largest $x_\star$) is always a minimum if $\mu>0$ and maximum if $\mu<0$. The black thick and dashed lines in the left panel separates between the regions with $V''_{eff} (x_\star)>0$ and $V''_{eff} (x_\star)<0$. The shaded region indicates the event horizon. The innermost extrema at $x=0$ are not shown.}}\label{FigVeffExtremalTimelike}
\end{figure}
Fig. \ref{FigVeffExtremalTimelike} encapsulates what is going on. We plot the ${\cal E}$-$x_\star$ contours of several values of $\ell$ for one of the BH solutions which appear in the previous sections. For a given $\ell$  there exists a curve (single branched or double branched) which may be crossed by the line of constant  ${\cal E}$ once or in three points and a degenerate possibility of two times exists as well. These points represent the minima or maxima of $V_{eff} (x)$ which alternate such that the largest is always a minimum if $\mu>0$ where $V_{eff} (x)$ is attractive. For the BH solutions with $\mu<0$ the exterior extremum is a maximum and the gravitational field is accordingly repulsive. The line ${\cal E}^2 (x_\star)$ of Eq. (\ref{LineVeffppEq0}) is the black thick line in the left panel of Fig. \ref{FigVeffExtremalTimelike}.

This plot shows quite an involved structure, but we will limit the present discussion to the trajectories which are always outside the event horizon, i.e. those which perform radial oscillations around the most exterior minimum of $V_{eff} (x)$, or make scattering trajectories with a minimal distance from the BH. Positive mass MHBHs will allow both kinds of trajectories for massive particles, but only the ``hyperbolic'' ones for light. Similarly, negative mass MHBHs will allow only ``hyperbolic'' trajectories. On the other hand, Fig.  \ref{FigVeffExtremalTimelike} demonstrates also that above a certain energy, all trajectories cross the event horizon.

An important special case is the case of circular orbits. The circular orbits exist if $V_{eff} (x_c)=V'_{eff} (x_c)=0$ is satisfied. These two equations may be solved explicitly such that for particles ($\epsilon=1$) we obtain the energy and angular momentum as a function of the orbital radius $x_c $
\be
{\cal E}^2_c = \frac{2 \left(x_c^4+1\right)^{5/2} f^2(x_c)}{x_c^6 \left( (3 x_c^4+2) f(x_c)+x_c^2 (p-x_c^2)\right)} \;\;\;\;\; , \;\;\;\;\;
\ell^2_c =-\frac{ x_c^4 \left(x_c^2 f(x_c)+p-x_c^2\right)}{\left(3 x_c^4+2\right) f(x_c)+x_c^2 \left(p-x_c^2\right)}
\label{CircularOrbitsLEparticles}
\ee
provided the two following additional conditions are also satisfied:
\be
(3 x_c^4+2) f(x_c)+x_c^2 (p-x_c^2)>0 \;\;\;\;\; , \;\;\;\;\;x_c^2 f(x_c)+p-x_c^2<0
\label{CircularOrbitsExistCond}
\ee
However, for photons ($\epsilon=0$) the two equations for ${\cal E}^2$ and $\ell^2$ become homogeneous, so only their ratio can be expressed in terms of $x_c$ and $x_c$ itself which we designate here $x_c^0$ should satisfy the additional condition
\be
(3 (x_c^0)^4+2) f(x_c^0)+(x_c^0)^2 (p-(x_c^0)^2)=0
\label{CircularOrbitsXClight}
\ee
As usual, these circular orbits are unstable.

The other light trajectories are obtained easily from the effective potential for photons, i.e.  Eq. (\ref{EffPotNeut}) with $\epsilon=0$. The main difference with respect to particles is that $V_{eff} (x)$ is asymptotically decreasing to the negative value of $-{\cal E}^2$ while its only minimum is inside the horizon. Thus, for photons there are no closed (``bound'') orbits outside the horizon. The general shape of $V_{eff} (x)$ is very similar to that of the left panel of Fig. \ref{VeffTimelike} which corresponds to open trajectories.

Finally we turn to the geodesics around the $\gamma<0$ MHBHs for either massive particles or light. The main new feature of these black holes is the appearance of curvature singularity at $r=r_s$ inside the horizon. On the other hand, outside the horizon the metric tensor in this case is quite similar to the $M>0$ ones for $\gamma>0$ and thus also the effective potential $V_{eff} (x)$. The particle trajectories are determined by an effective potential which looks like that in Fig. \ref{VeffTimelike} differing mainly in that $V_{eff} (x)$ does not tend to a positive constant at the origin, but diverges as $r\rightarrow r_s$. Outside the horizon, the pattern is conventional: the timelike geodesics are either closed for ${\cal E}<1$, or open for ${\cal E}>1$. The geodesics of this second kind can be either ``hyperbolic'' or absorbed by the BH after coming from spatial infinity with small enough angular momentum.

For null geodesics we  also have a similar pattern as for $\gamma>0$: either unstable circular orbits, or more typically, open trajectories.

\section{Trajectories of Charged Particles}  \label{GeodesicsCharged}
\setcounter{equation}{0}
Charged particles ``feel'' also the magnetic field of the Horndeski magnetic BH, so they do not follow geodesics, but different trajectories which are described by solutions of the equations of motion which are derived from the general Lagrangian (\ref{GeodEqsMagLagrangian}) for $q \ne 0$.

It is obvious that the additional magnetic force does not allow planar trajectories. However, since the system is still spherically symmetric, a conserved angular momentum still exists with the following components:
\bea
J_1 &=&  -m r^2 \left(\dot{\theta}\sin\phi+\dot{\phi}\sin\theta\cos\theta\cos\phi\right)+
qP\sin\theta\cos\phi  \nonumber \\
J_2 &=&   m r^2 \left(\dot{\theta}\cos\phi-\dot{\phi}\sin\theta\cos\theta\sin\phi\right)+
qP\sin\theta\sin\phi
\\
J_3 &=& m r^2 \dot{\phi}\sin^2\theta +qP\cos\theta \nonumber
\label{ConservedAngularMomentum}
\eea
Since the three components are conserved, the axes may be rotated such that  $J_1 = J_2 =0$ and only $J_3$ is non-zero. From this follows the  constraint $\dot{\theta} =0$ and also the constant value of the polar angular coordinate is found to be $\cos \theta_0 = qP/J_3$.

 The equations of motion turn out to be a slight modification of Eqs. (\ref{GeodEqsMagIntegrated}) for the same variables $x(\tau)$ and $\phi(\tau)$:
\be
x^2 \dot{\phi}  = {\cal J} \,\,\, ; \,\,\,\,\,\, \dot{x}^2+f(x)\left( \epsilon +\frac{{\cal J}^2 \sin^2\theta_0}{x^2} \right) - \frac{{\cal E}^2}{a^2(x)} = 0
\label{GeodEqsCharged}\ee
where ${\cal J}$ is the rescaled total angular momentum. So formally, one may obtain the radial component of the motion of charged particles, $x(\tau)$ by replacing $\ell \mapsto {\cal J} \sin\theta_0$ in the solutions of Sec. \ref{GeodesicsNeut}. However, the solutions are obviously significantly  different physically, since they are not planar, but confined to a conical-like surface whose apex is at $r=0$. For example, the circular orbits discussed at the end of Sec.  \ref{GeodesicsNeut} translate simply to the case of $q \ne 0$. However these circular orbits do not have their center at $r=0$, and $x_c$ is not the dimensionless coordinate radius, but the radial distance from the origin.

\section{Light Deflection}  \label{LightDeflection}


Frequently, it is the shape of particle trajectories  around BHs which is more interesting than the time dependence since it it usually used for calculating observable quantities like light deflection and perihelion precession. The shape, $x(\phi)$ is determined by transforming the second (radial) equation of (\ref{GeodEqsCharged}) for  $x(\tau)$ into an equation for $x(\phi)$ using the first equation of (\ref{GeodEqsCharged}). This gives:
\be
(x')^2+U^{(q)}_{eff} (x)=0 \,\,\, , \,\,\,\,\,\, U^{(q)}_{eff} (x)=\frac{x^4}{{\cal J}^2}V^{(q)}_{eff} (x)= f(x)\left( \frac{\epsilon ~x^4}{{\cal J}^2} +\sin^2\theta_0  ~x^2 \right) - \frac{{\cal E}^2}{{\cal J}^2}\frac{x^4}{a^2(x)}
\label{GeodEqsChargedPhi}\ee
where $x' = dx/d\phi$ and $V^{(q)}_{eff} (x)$ is the effective potential for charged particles appearing in (\ref{GeodEqsCharged}).

Following the standard procedure we will make a further change of variables to obtain an equation for $s(\phi)=1/x(\phi)$:
\be
(s')^2+W^{(q)}_{eff} (s) = 0\,\,\, , \,\,\,\,\,\, W^{(q)}_{eff} (s)=V^{(q)}_{eff} (1/s)/{\cal J}^2 = f(1/s)\left( \frac{\epsilon}{{\cal J}^2} +\sin^2\theta_0  ~s^2 \right) - \frac{{\cal E}^2/{\cal J}^2}{ a^2(1/s)}
\label{OrbitsEqsNeut}\ee

Here we will study only null geodesics of neutral particles (i.e. light rays), so we will use only $U_{eff}(x)=U^{(q=0)}_{eff}(x)$ and $W_{eff}(s)=W^{(q=0)}_{eff}(s)$ with $\theta=\pi/2$ and ${\cal J} = \ell$ and later take $\epsilon = 0$.

In order to calculate the deflection of a neutral particle trajectory by the MHBH, we need to calculate the angle $2\Delta\phi$ between the outgoing asymptotical direction of a ``hyperbolic'' trajectory and its asymptotic incoming direction. The deflection angle $\psi$ is just
given by $\psi = 2\Delta\phi - \pi$.

 The angle $2\Delta\phi$ is calculated simply as twice the angle between the point of closest approach (``periastron'') and the asymptotic outgoing (or incoming) direction:
\be
\Delta\phi = \int_0^{s_p} \frac{ds}{\sqrt{-W^{(q=0)}_{eff}(s)}} = \int_0^{s_p} \frac{ds}{\sqrt{f(1/s_p)\left( \epsilon / \ell^2 + s_p^2 \right)\frac{ a^2(1/s_p)}{ a^2(1/s)}-f(1/s)\left( \epsilon / \ell^2 + s^2 \right)}}
\label{DeflectionGeneral}\ee
where $s_p = 1/x_p$ corresponds to the point of closest approach located at the coordinate distance $x(\phi=0)=x_p$. We also expressed the energy parameter ${\cal E}$ in terms of the other relevant parameters by using the equation $W_{eff} (s_p)=0$.

In order to calculate $\Delta\phi$ we have to substitute in (\ref{DeflectionGeneral}) the explicit expressions for the MHBH metric functions given in Eqs (\ref{Sola}) and (\ref{Solfr}) (or rather their dimensionless versions - see (\ref{Solfx})). However, it seems that expressing the integral expression for $\Delta\phi$ in terms of elementary functions is not possible, not even for massless particles for which the right-hand-side of (\ref{DeflectionGeneral}) simplifies considerably.

What is found to be possible is expressing the deflection angle for massless particles (i.e. light; $\epsilon=0$) as a series expansion in powers of the dimensionless inverse ``periastron'' distance $s_p = 1/x_p$ as can also be done for the \textbf{S} solution.

An elementary but lengthy calculation gives the following expansion up to third order for the deflection angle for light:
\be
\psi(\mu,p,s_p)=2\mu s_p +\left[\left(\frac{15 \pi
   }{16}-1\right) \mu ^2-\frac{3 \pi  p}{4}\right] s_p^2+
   \left[\left(\frac{61}{12}-\frac{15 \pi }{16}\right) \mu ^3-\left(7-\frac{3 \pi }{4}\right) \mu  p\right] s_p^3 +\cdots
\label{DeflectionExpansion}\ee
we notice that the $p\rightarrow 0$ limit yields the usual schwarzschild result.

\begin{figure}[t!!]
\begin{center}
{\includegraphics[width=7.75cm, angle = -00]{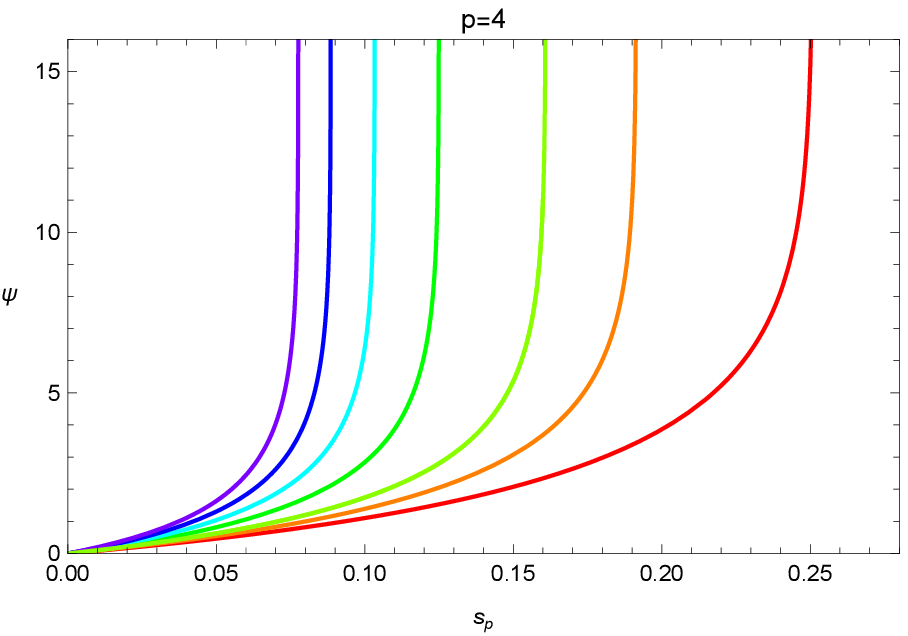}}\hspace{5mm}
{\includegraphics[width=7.75cm, angle = -00]{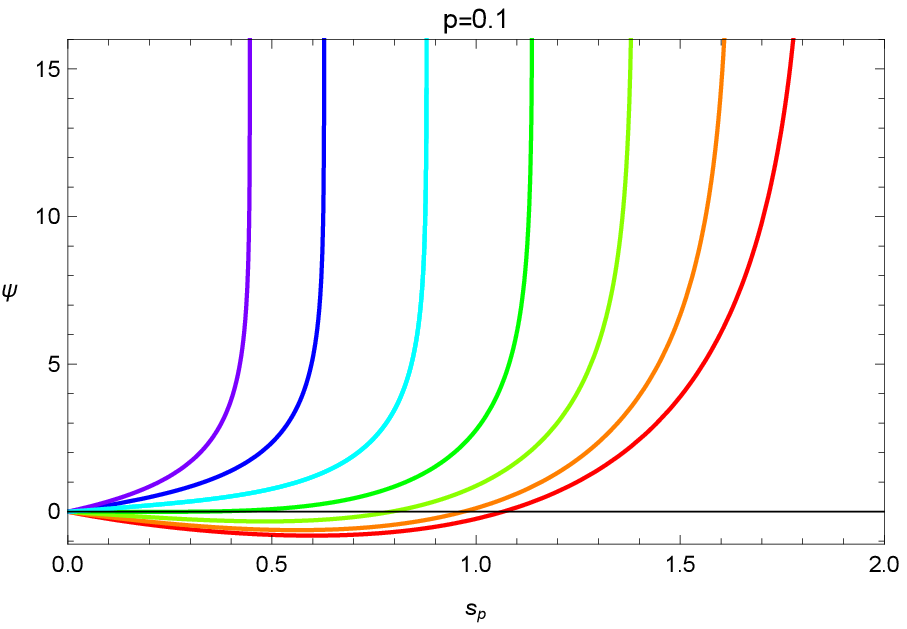}}
\end{center}
\caption{{\small The dependence on the inverse of the ``periastron'' coordinate  of light deflection angle from  typical MHBHs with large (left) and small (right) magnetic charges and several values of mass. $\gamma>0$. The parameters are: Left: $p=4$ , $\mu=3.988(\textit{extremal BH}),\; 4.5,\; 5,\; 6,\; 7,\; 8,\; 9$ ; Right: $p=0.1$ , $\mu=-1.390 (\textit{extremal}),\; -1,\; -0.5,\; 0,\; 0.5,\; 1,\; 1.5$. Notice that for negative mass the deflection angles are negative for large ``periastron'' (small $s_p$).}}\label{FigLightDeflection-MagFixed-p}
\end{figure}

It is quite simple to integrate Eq. (\ref{DeflectionGeneral}) numerically (also for $\epsilon=0$) and to present sections of the three-variable function $\psi(\mu,p,s_p)$. Fig. \ref{FigLightDeflection-MagFixed-p} depicts a representative portion of this function for $\gamma>0$. The diverging behavior of the deflection angle is another aspect of the existence of photon sphere around the MHBHs. For small angles the behavior is very well approximated by  the expansion (\ref{DeflectionExpansion}). The negative deflection angles for the negative mass solutions are of course a result of the gravitational repulsion in this case.

As before, the results for light deflection around $\gamma<0$ MHBHs are very similar to the case of $\gamma>0$. We also found that the power expansion (\ref{DeflectionExpansion})  for the deflection angle does not depend on the sign of  $\gamma$ up to third order and is valid as is for $\gamma<0$ as well. A difference between the two signs appears only in the fourth order. The only significant difference is for small $p$ where there exist negative mass BHs for $\gamma>0$, but not for $\gamma<0$. This means that the light deflection by $\gamma<0$ MHBHs is depicted  to a very good approximation by the left panel of Fig. \ref{FigLightDeflection-MagFixed-p} only.

\section{Comparison with the Electric Black Holes}  \label{ElectricBHs}
\setcounter{equation}{0}
The electric type of the vector-tensor Horndeski black holes (the EHBH) were constructed (numerically) and discussed already from several aspects  \cite{muller,BalakinEtAl2008,YB+YV2020}. Still, it is of interest to compare them on the same basis with their magnetic counterparts. In this section we will present briefly the EHBHs with adaptations to this context and proceed further in studying the general structure of the space of solutions and some concrete aspects like the pattern of horizons and the related BH temperature.

The field equations of electrically charged (charge $Q$) BH of the same theory are
\be
\label{EqaEl}
\frac{r a'}{a}-\frac{\gamma  \kappa^2  Q^2}{\left(2 \gamma\kappa  (1-f)+r^2\right)^2}=0
\ee
\be
\label{EqfEl}
r f'+f-1+\frac{\kappa  Q^2}{2 \left(2 \gamma\kappa  (1-f)+r^2\right)}=0
\ee
with the additional ``Maxwell'' equation for the electrostatic potential ${\cal V}(r)$ (or field $E=-{\cal V}'/a$)
\be\label{EqMaxwellEl}
-\frac{\left(r^2 +2 \gamma\kappa(1-f)\right){\cal V}' }{a}=Q
\ee
Unlike the magnetic equations, these are non-linear and like previous researchers we were unable to find explicit analytic solutions even in the very simple form which is obtained in terms of the mass function $M(r)$ defined by $f(r)=1-2M(r)/r$:
\be
\label{EqsMfEl}
\frac{a'}{a}-\frac{\gamma  \kappa^2  Q^2 r}{\left(4 \gamma\kappa  M+r^3\right)^2}=0
\;\;\; , \;\;\;
M'-\frac{\kappa  Q^2 r}{4 \left(4 \gamma\kappa  M +r^3\right)}=0
\ee
If we rescale the variables by the natural length scale of the system $\ell_{el} = |\gamma|^{1/4}\sqrt{\kappa Q}$ such that $x=r/\ell_{el}$, we obtain for $a(x)$ and $m(x)=4|\gamma|\kappa M(\ell_{el} x)/\ell_{el}^3$ the simple equations
\be
\label{EqsMfElRescaled}
a'/a=x/\left(\pm m+x^3\right)^2
\;\;\; , \;\;\;
(\pm m+x^3)m'=x
\ee
where $\pm$ correspond to both possible signs of $\gamma$. Indeed, the explicit solution has avoided us but still, solving the $m(x)$--equation numerically is almost trivial and from the result for $m(x)$, it is equally easy to solve the first equation  for $a(x)$. We notice also that the rescaled system (\ref{EqsMfElRescaled}) contains no free parameters and is therefore ``universal'': the solutions $m(x), a(x)$ contain all the information about the EHBH solutions and their features which are obtained by simple scaling. Recall that this scaling symmetry is broken in the magnetic case where there are no ``universal'' solutions as can be seen from the magnetic analogs of Eqs. (\ref{EqsMfElRescaled}). Notice also that if one defines a rescaled dimensionless field $\varepsilon=|\gamma|^{1/2}\kappa E$, it may be expressed in terms of the mass function simply as $\varepsilon=m'$. However, one should note that the metric functions themselves are not universal. For example, the metric function $f(r)$ is written in terms of the rescaled dimensionless variables as $f(x)=1-qm(x)/x$ where $q=Q/2|\gamma|^{1/2}$ is the charge parameter defined in analogy to the magnetic parameter $p$. Without loss of generality we assume $Q>0$.

\begin{figure}[b!!]
\begin{center}
{\includegraphics[width=7.0cm, angle = -00]{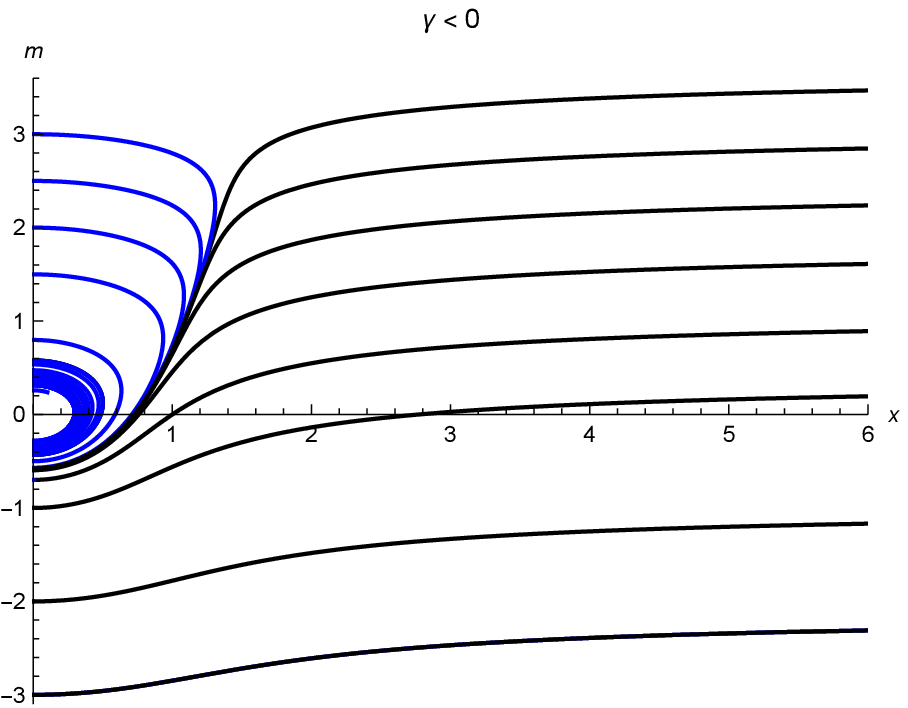}}\hspace{1cm}
{\includegraphics[width=7.0cm, angle = -00]{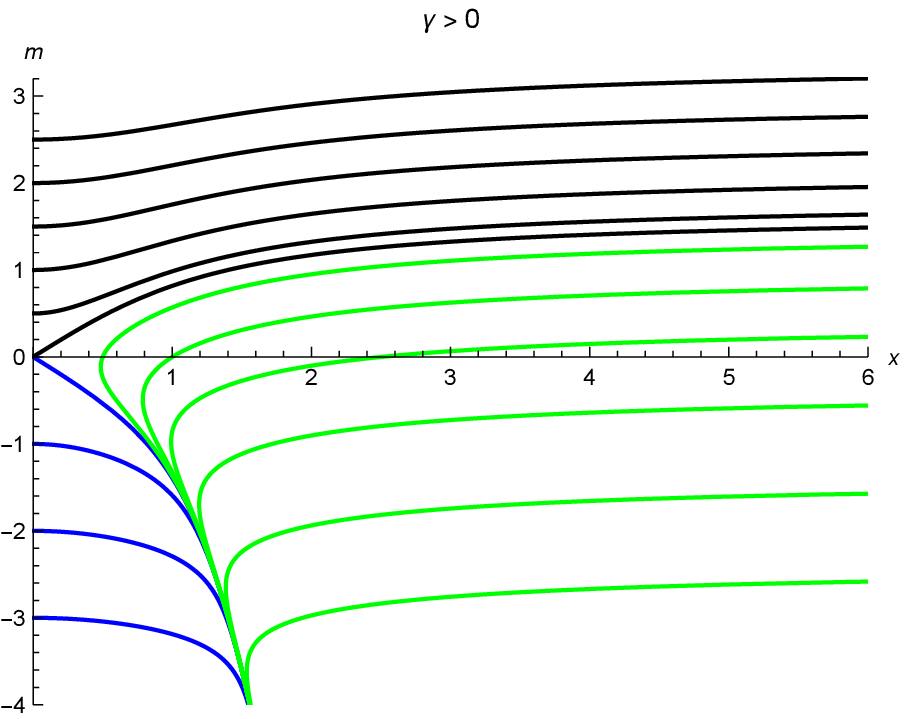}}
\end{center}
\caption{{\small All possible solutions of Eq. (\ref{EqsMfElRescaled}) for $m(x)$ with both signs of $\gamma$. The $\gamma>0$ BH solutions (right panel) correspond to the family of the upper (black) curves and to those solutions of the lower-right (green) family with positive $m(\infty)$. The $\gamma<0$ BH solutions (left panel) correspond to the (black) curves of the family of solutions which are defined for all $x\geq 0$ which also have a positive $m(\infty)$.
}} \label{FigMStreamlines}
\end{figure}

The  solutions $m(x)$ and $a(x)$ which correspond to asymptotically flat EHBHs may be characterized by the mass parameter $m=m(\infty)$, but in practice it may be more convenient to do it either by the value of $m(0)$ when $m(x)$ is regular  for all $x\geq 0$, or at least defined for $x=0$, or if this is not the case, by another boundary condition like the coordinate $x_1$ for which $m(x_1)=0$, or the horizon $x_H$ satisfying $x_H = qm(x_H)$, etc. The second integration constant related to $a(x)$ is determined by the asymptotic condition $a(x)\rightarrow 1$ at infinity. Fig. \ref{FigMStreamlines} presents all the possible solutions of the mass function $m(x)$ for both signs of $\gamma$ which determine the two metric functions. For $\gamma>0$ we can identify 3 types of solutions. The first are solutions which are regular for all $x\geq 0$. These are the upper (online black) curves in the figure and they correspond obviously to BH solutions. These solutions have a mass which is bounded from below such that $m(\infty)\geq 1.6543$ where the lowest mass curve is also characterized by $m(0)=0$. This is actually an intermediate solution between this type and the second type of solutions represented by the (green) partially double-valued curves of the lower right hand side of the plot. These $m(x)$ solutions are evidently singular at $x=x_s$ - the point of minimal $x$ value. As long as $m(\infty)>0$ (but of course $m(\infty)<1.6543$), these solutions may correspond to BHs with a singularity on the spherical surface $x=x_s$. The lower branches of these solutions and those curves which lie entirely below the $x$-axis are unphysical. The third type are the negative $m(x)$ (blue) curves with diverging mass function. They diverge like $m(x)\sim -x^3$ and obviously cannot correspond to BHs.

These EHBH solutions have the following power series expansion near the origin:
\bea \nonumber
m(x)=m_{0}+ \frac{x^2}{2m_{0}}- \frac{x^4}{8m_{0}^3}- \frac{x^5}{5m_{0}^2} +\frac{x^6}{16m_{0}^5}+ \frac{6x^7}{35m_{0}^4}+ ... \;\;\;\; ,\;\;  m_{0}>0 \\
m(x)=x - \frac{x^3}{4}+ \frac{3x^5}{32} - \frac{9x^7}{256}+ ... \;\;\;\; ,\;\; m_{0}=0
\label{EHBHpowerSeries}
\eea
For the singular solutions we get the behavior near the singularity as an expansion of the inverse function $x(m)$ around its minimum at $m_0 = -x_0^3$:
\bea
x(m)=x_0+\frac{(m+x_0^3 )^2}{2x_0} +\frac{(m+x_0^3 )^3}{2}+\frac{(3x_0^4-1)(m+x_0^3 )^4}{8x_0^3} + ...
\label{EHBHpowerSeriesInverse}
\eea
Asymptotically, all these solutions have the usual RN behavior.

For $\gamma<0$ there are only 2 different types of solutions: the first is solutions which exist only in a finite interval of the radial coordinate and obviously cannot correspond to BHs. These are the blue curves of Fig. \ref{FigMStreamlines}. The second are monotonically increasing solutions defined for all $x\geq 0$ similar to those of the $\gamma>0$ case. However, unlike those, the mass function of the $\gamma<0$ solutions is never positive definite. It is either negative for all $x\geq 0$, or starts with a negative value of $m(0)$ and changes sign so it becomes positive with a finite mass parameter $m=m(\infty)$. This type of solutions with $m(\infty)>0$ may correspond to BH solutions. They exist for a limited interval of $m(0)$ values, namely $-1.2276<m(0)<-0.5791$. However, the mass $m(\infty)$ is not bounded from above.

The behavior of the EHBH mass function $m(x)$ near the origin is given now by
\bea
m(x)=m_{0}- \frac{x^2}{2m_{0}}- \frac{x^4}{8m_{0}^3}- \frac{x^5}{5m_{0}^2} -\frac{x^6}{16m_{0}^5}- \frac{6x^7}{35m_{0}^4}+ ... \;\;\;\; ,\;\;  m_{0}<0
\label{EHBHpowerSeriesGamNeg}
\eea

Next we discuss the metric functions $g_{rr}(x)=-1/f(x)$ and $g_{00}(x)=f(x)a^2 (x)$ of these 3 types of EHBH solutions, skipping $a(x)$ which is obtained very easily when $m(x)$ is at hand. As mentioned already, the metric functions are not universal, but depend also on the dimensionless charge parameter $q$ since $f(x)=1-qm(x)/x$. Therefore, the charge parameter $q$ introduces an additional structure to the space of solutions; most importantly to the BH horizon pattern. The existence of a horizon depends on the existence of a solution to the equation $x=qm(x)$ while we notice that $m(x)$ is always monotonically increasing and bounded for the BH candidate solutions. Thus, for the positive definite $m(x)$ solutions of $\gamma>0$ there will be always a single solution to the equation $x=qm(x)$ for any finite value of the charge parameter $q$. That means a single horizon for any mass in the range $m(\infty)> 1.6543$. These BHs will be S-like. For the other two kinds of mass curves - one for each sign of $\gamma$, the equation $x=qm(x)$ may have two solutions, or one, or none at all according to the values of $q$ and $m(\infty)$. This is the typical situation for RN-like BHs which exist for the lower mass range of $0< m(\infty)< 1.6543$ for the $\gamma>0$ EHBHs and all $m(\infty)>0$ for the $\gamma<0$ EHBHs. So the bottom line is that all $\gamma<0$ EHBHs are RN-like, while for $\gamma>0$ the mass parameter determines uniquely whether the BH is S-like or RN-like.
\begin{figure}[b!!]
\begin{center}
{\includegraphics[width=8cm, angle = -00]{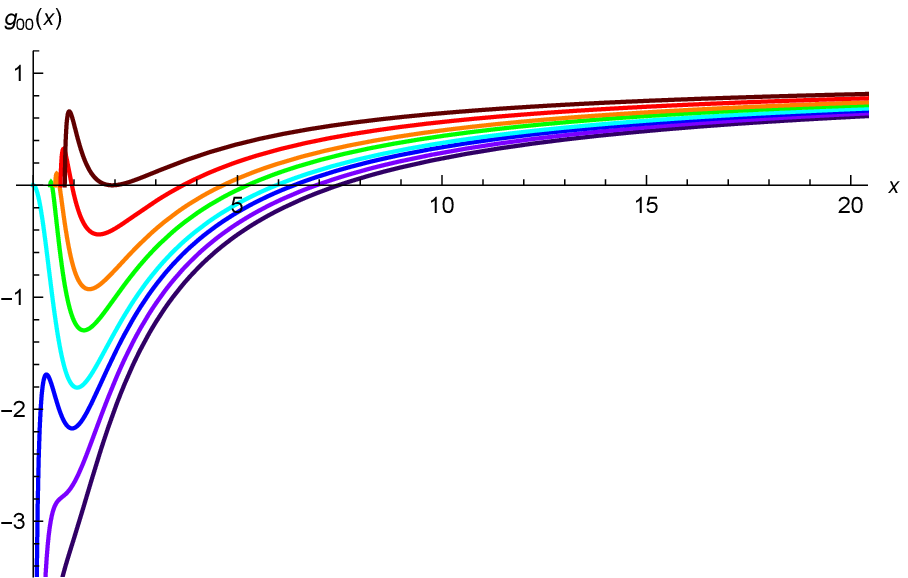}}
{\includegraphics[width=8cm, angle = -00]{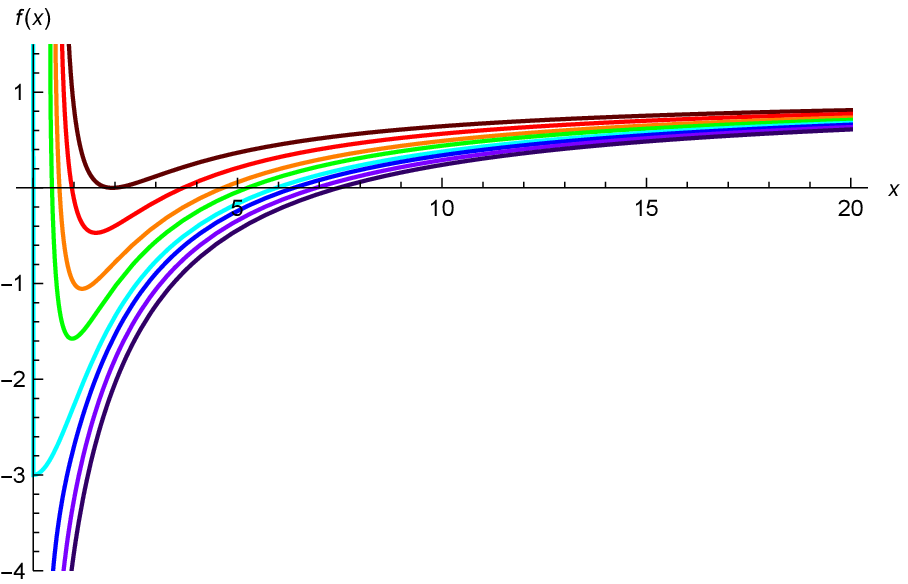}}
\end{center}
\caption{{\small Profiles of  $g_{00}(x)$ and of $f(x)$ for EHBHs with $\gamma>0$, $q=4$
and several values of the  mass parameter: $\mu=qm(\infty)=3.962$ (extremal)$,\;4.75,\; 5.5,\;6,\; 6.617$ (intermediate, $m(\infty)=1.654$),$\;7,\; 7.5,\;8$.}} \label{FigProfilesElectricGamPos}
\end{figure}

\begin{figure}[t!!]
\begin{center}
{\includegraphics[width=8cm, angle = -00]{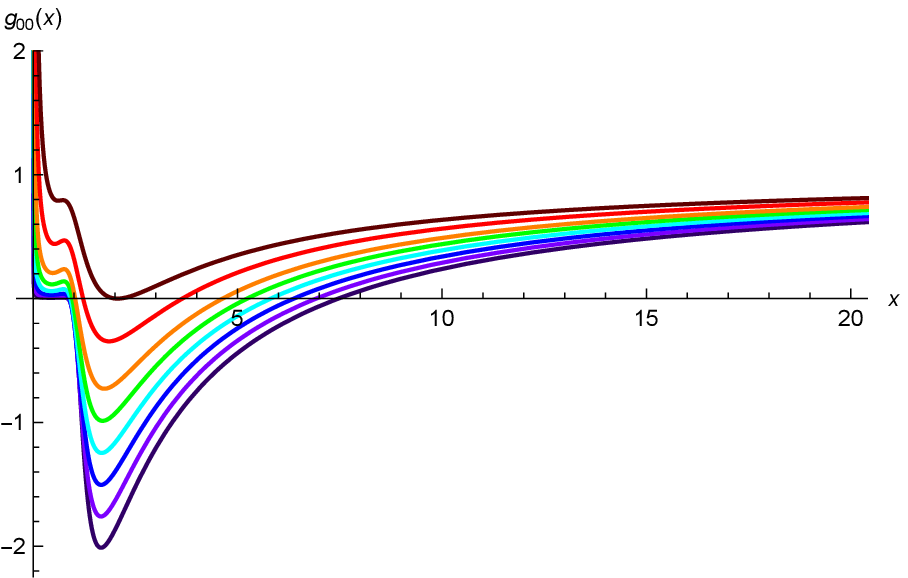}}
{\includegraphics[width=8cm, angle = -00]{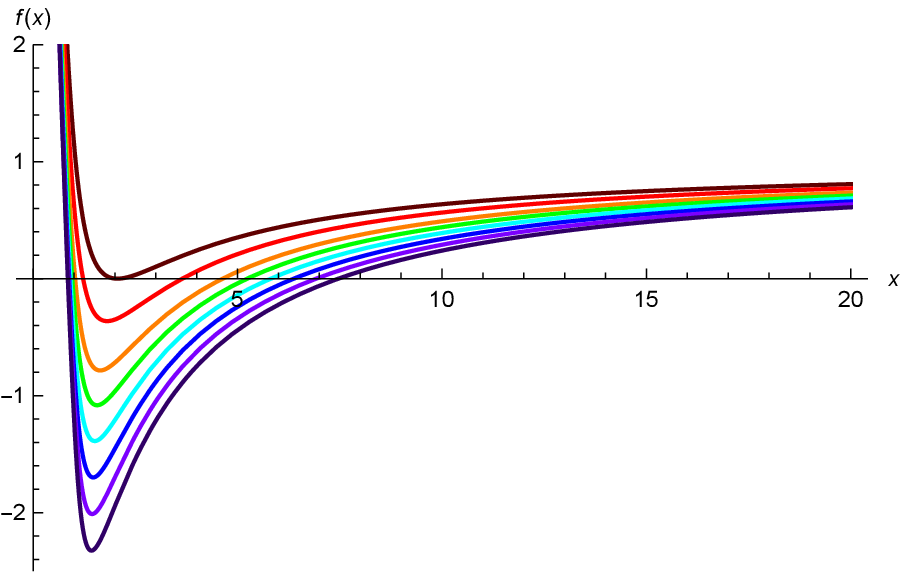}}
\end{center}
\caption{{\small Profiles of  $g_{00}(x)$ and of $f(x)$ for EHBHs with $\gamma<0$, $q=4$
and several values of the  mass parameter starting from the extremal solution: $\mu=qm(\infty)=4.037,\;4.75,\; 5.5,\; 6,\;6.5,\; 7,\;7.5,\; 8$.}} \label{FigProfilesElectricGamNeg}
\end{figure}

The metric functions of these 3 types of EHBH solutions are shown in Figs. \ref{FigProfilesElectricGamPos} (for $\gamma>0$) and \ref{FigProfilesElectricGamNeg} (for $\gamma<0$) in which they are plotted for the representative value $q=4$. We chose this specific value in analogy with the magnetic case it order to make the comparison between the electric and magnetic cases as clear and direct as possible. For this end we also characterize the solutions by the mass parameter $\mu=2M/\ell_{el} = qm(\infty)$ which is defined  in analogy with the magnetic case. We can indeed notice (as expected) that the profiles of the same $\mu$ and $q=p$ have (when exist) very similar behavior far enough outside the horizon. Few examples are obvious by comparing the pairs of Figs.  \ref{FigProfilesElectricGamPos},\ref{FigProfilesElectricGamNeg}  with Figs. \ref{FigProfiles},\ref{FigProfilesNegGam} respectively. We will now summarize the main characteristics of the EHBHs for both signs of $\gamma$ in addition to what we have already seen.

{\boldmath $\gamma$} $\mathbf{>0}.$
First we notice from Fig. \ref{FigProfilesElectricGamPos} that unlike the $\gamma>0$ magnetic solutions where $f(0)$ is always zero and $g_{00}(x)$ has mostly RN-like shape (upper part of Fig. \ref{FigProfiles}), the metric function $f(x)$ of the $\gamma>0$ electric solutions has RN-like behavior for small values of the mass parameter and ``Schwarzschild-like'' behavior for large mass values. The negative mass MHBHs do not have an electric analogue. For a given $q$ as in Fig. \ref{FigProfilesElectricGamPos}, the BH mass is bounded from below just as in the ordinary RN solutions. In the specific case of $q=4$ that we plot, the mass bound is $\mu=qm(\infty)=3.9622$ which corresponds to the extremal BH with a degenerate horizon. At the ``boundary'' between the RN-like and S-like solutions, there is a single special intermediate solution, regular for all $x\geq 0$, with universal $m(\infty)=1.6543$  that corresponds to the mass profile with $m(0)=0$ and $m'(0)=1$ which also results  $f(0)=1-q$ and $a(0)=0$. Therefore, we have in the figure for this intermediate solution  $f(0)=-3$ and $g_{00}(0)=0$. Although the metric components of this solution are regular at the origin, it is still a point of curvature singularity as can be suspected from the fact that $a(0)=0$ and seen explicitly from the curvature invariants. Compare the similar solution which exists in the magnetic case for $\gamma<0$. See e.g. Fig. \ref{FigProfilesNegGam}.

As mentioned already, the S-like EHBHs can support any finite value of the charge parameter $q$.
 However, the smaller mass RN-like solutions need a charge parameter above a certain minimum $q_{min}(m)$ which depends on its mass. It is quite easy to see that there is a minimal value of $q_{min}(m)$ which is obtained from the slope of the intermediate solution curve at $x=0$: $q_{min}(m)>m_{int}'(0)=1$. In other words, for $q<1$ the $\gamma>0$ EHBHs will be all S-like. Note however, that the relation between the actual mass, charge and horizon size,  $M$ , $Q$ and  $r_H$, is not directly reflected from the relation between the corresponding rescaled parameters, since the rescaling involves the charge $Q$. We will turn to the actual relations shortly.

Another new feature of the RN-like EHBHs is the appearance of a curvature singularity which these solutions ``inherit'' from the $m(x)$ singularity at $x=x_s$, so these solutions are defined only for $x>x_s$, with  $x_s$ decreasing with $m(\infty)$ as can be seen from the $f(x)$ profiles of Fig. \ref{FigProfilesElectricGamPos}, or from the (green) mass profiles of Fig. \ref{FigMStreamlines}. Notice also that this is a somewhat uncommon kind of singularity: as is obvious from Fig. \ref{FigMStreamlines} (or from inspection of Eq. (\ref{EqsMfElRescaled})), it is the derivative $m'(x)$ which is singular, while the mass function $m(x_s)$ itself is finite. Accordingly, the nature of the singularity of $f(x)$ is similar:  its derivative will diverge as $x\downarrow x_s$, but $f(x_s)$ will be finite. This feature is not seen in Fig. \ref{FigProfilesElectricGamPos} since the value of $f(x_s)$ is too high for the figure frame. We have also checked that the Ricci scalar diverges at the singular point.

{\boldmath $\gamma$} $\mathbf{<0}.$
The $\gamma < 0$ EHBH solutions are all RN-like and they all differ from their $\gamma > 0$ counterparts in that they are all singular only at the origin. Their general behavior is more similar to that of the ordinary RN solutions: their mass is bounded from below (by $\mu=qm(\infty)=4.0372$ for the present $q=4$) but unbounded from above. For a given mass, EHBHs will exist only above a minimal $q_{min}(m)$ as for $\gamma>0$. We also notice that as the mass increases $a(0)$ becomes vanishingly small inside the event horizon and so does $g_{00}(x)$, but $f(x)$ keeps its RN-like behavior.

\begin{figure}[t!!!]
\begin{center}
{\includegraphics[width=8cm, angle = -00]{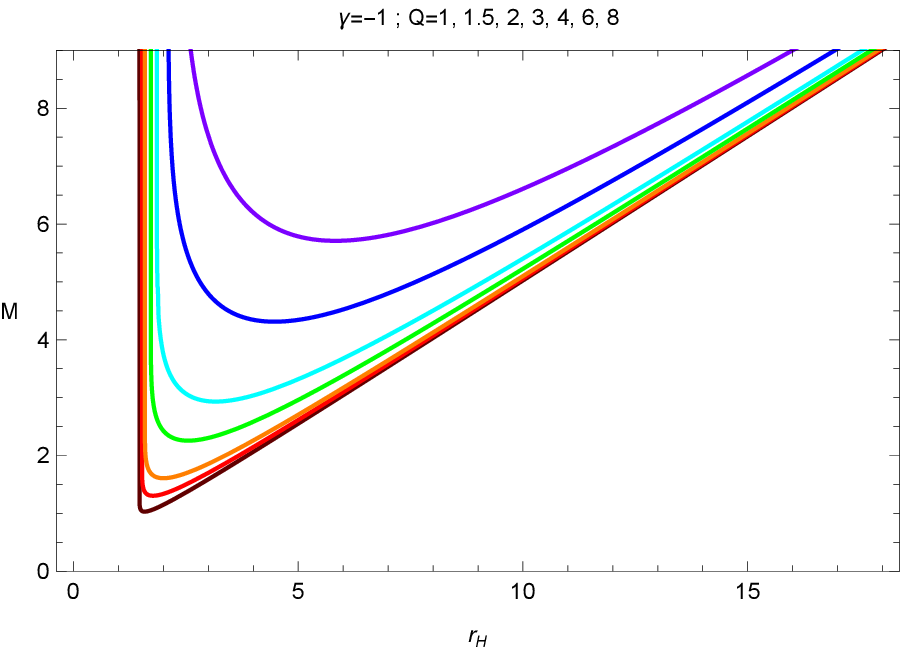}}
{\includegraphics[width=8cm, angle = -00]{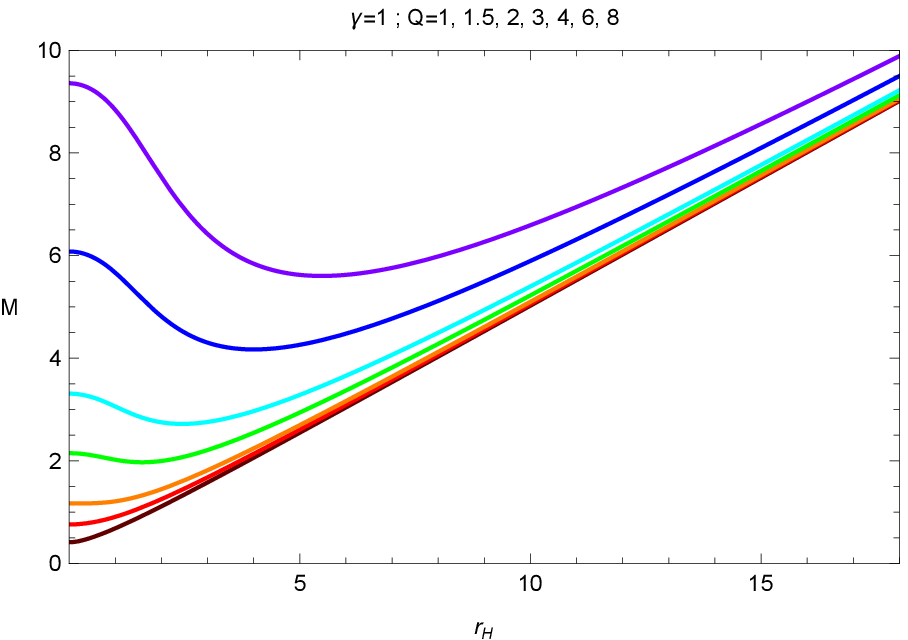}}
\end{center}
\caption{{\small The mass $M$ of EHBHs as a function of the horizon size for several values of the charge $Q$. Left:  $\gamma<0$ ; Right:  $\gamma>0$. }} \label{FigElBHMvsrH}
\end{figure}


An additional angle of comparison between the electric and the magnetic case is to compare the asymptotic behavior of the metric components. For the MHBHs it was obtained in Eq. (\ref{Asymptf}) directly from the explicit solution. In the electric case we have to use directly the field equations to find the asymptotic behavior of $f(r)$:
\be
\label{AsymptfElectric}
f(r)=1-\frac{2M}{r}+\frac{\kappa Q^2}{2 r^2}-\frac{\gamma\kappa^2 Q^2 M}{2 r^5}+\frac{\gamma\kappa^3 Q^4}{10 r^6}+...
\ee
The behavior is identical in both cases up to the terms of second order, but beyond that, noticeable differences appear which again reflect the explicitly broken duality symmetry of this theory.\\

Finally we will obtain for both signs of $\gamma$ the BH mass $M$ and temperature $T$ vs. the actual charge $Q$ and the horizon coordinate $r_H$ as we did for the MHBHs. Regarding the mass, since we do not have explicit analytic solutions, we cannot present the full dependence by a function like $M(\gamma,Q,r_H)$ in analogy with the magnetic case (see (\ref{BHMass})), but we will be content with plotting several sections of $M$ vs. $r_H$ for certain values of $Q$. Fig. \ref{FigElBHMvsrH} shows these sections for both signs of $\gamma$. The $M(r_H)$ curves are noticeably different from their magnetic counterparts for $\gamma>0$: $M(r_H)$ does not diverge as $r_H\rightarrow 0$; on the contrary. Moreover, if we plot the rescaled quantities for $\gamma>0$ we will see that all curves merge at the same point at $x_H =0$ with the (universal) mass parameter of the intermediate solution $m(\infty)=1.6543$. This mass parameter is the minimal for all  $q\leq 1$ EHBHs ($Q\leq 2$ in Fig. \ref{FigElBHMvsrH}) which are all S-like. For $q>1$ ($Q> 2$ in Fig. \ref{FigElBHMvsrH}), the curves develop a lower minimum as happens for the RN solutions. The  $M(r_H)$ curves with $\gamma<0$ also differ significantly from those of the magnetic case. Unlike those, the inner horizon is not bounded by the origin, but by a certain value $r_{Hmin}$ which depends on $Q$. The mass curve diverges for $r_H \downarrow r_{Hmin}$ and also as usual for $r_H \rightarrow \infty$. Also as usual, for any fixed $Q$ there is a minimal mass which corresponds to the extremal BH solution which has a maximal charge to mass ratio.

\begin{figure}[t!!!]
\begin{center}
{\includegraphics[width=8cm, angle = -00]{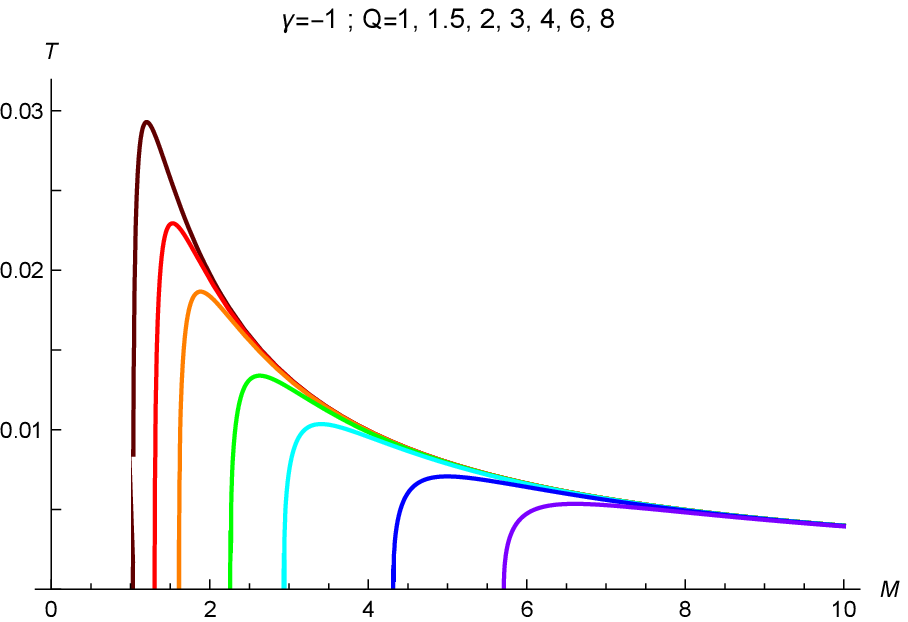}}
{\includegraphics[width=8.25cm, angle = -00]{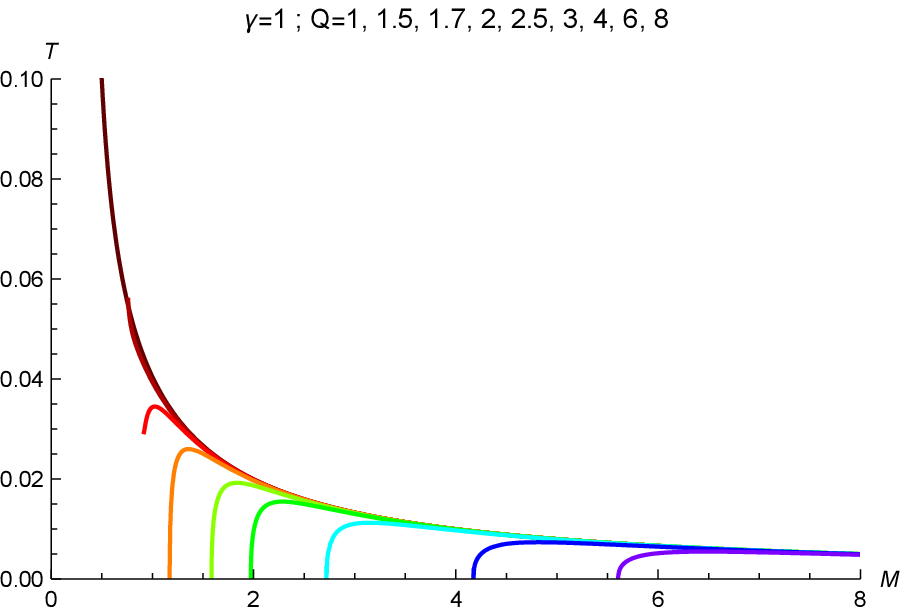}}
\end{center}
\caption{{\small The temperature $T$ of EHBHs as a function of the mass $M$ for several values of the charge $Q$. Left:  $\gamma<0$ ; Right:  $\gamma>0$. For $Q\ge 2$ the curves for both signs of $\gamma$ are quite similar in shape and value. For $Q< 2$ there is a sharp difference. Notice that the $\gamma>0$ curve for $Q=1.5$ almost merges with the lower part of the $Q=1$ curve. }} \label{FigElBHTvsM}
\end{figure}

For the temperature we can obtain an ``almost analytic'' expression for the $r_H$ and $Q$ dependence using Eq. (\ref{EqfEl}):
\be
T(\gamma, Q , r_H)= \frac{a(r_H)f'(r_H)}{4\pi}= \frac{a(r_H)(r_H^2+2\gamma\kappa-\kappa Q^2 /2)}{4 \pi  r_H (r_H^2+2\gamma\kappa)},
 \label{BHTemperatureElectric}
\ee
but with $a(r_H)$ obtained from the numerical analysis. Thus we will present the mass dependence of the BH temperature for several values of the charge $Q$ in Fig. \ref{FigElBHTvsM}. Since all the $\gamma<0$ BHs are RN-like, the $T(M)$ curves take the ordinary shape. Notice that on first sight, for $\gamma<0$ the temperature may diverge as $r_H \rightarrow \sqrt{-2\gamma\kappa}$. However this cannot happen since it is obvious that $r_{ext}$ for which the numerator of Eq. (\ref{BHTemperatureElectric}) vanishes, is always larger than $\sqrt{-2\gamma\kappa}$.

The $\gamma>0$ curves which cross the $M$-axis, i.e. start with a minimal mass and $T=0$, are indeed quite similar to the $\gamma<0$ ones, but there is a difference:  these curves correspond only in their lower mass part to RN-like BHs, but they continuously deform and change to S-like BHs as seen in Fig. \ref{FigProfilesElectricGamPos}. These curves correspond to soutions with $q\geq 1$. The other kind of $\gamma>0$ curves with $q< 1$ (which are all S-like as mentioned already) do not cross the $M$-axis, although some of the larger $q$-values may start rising from a point close to the $M$-axis and then decrease further with $M$. The other curves of this branch decrease monotonically with $M$ starting from a point of non-zero minimal mass and maximal temperature. The reason of this is of course the charge $Q$ which forces the EHBH mass away from zero.



\section{Particle Trajectories and Light Deflection -- Electric Case}  \label{LightDeflectionEl}
\setcounter{equation}{0}

The trajectories of test particles around the electric BH are determined by the Lagrangian
\be
L=-m\sqrt{a^2 f(r)\dot{t}^2-\dot{r}^2/f(r)-r^2(\dot{\theta}^2+\sin^2\theta~\dot{\phi}^2)}-q' {\cal V}(r)\dot{t}
\label{GeodEqsElectricLagrangian}\ee
where $q'$ is the test charge (we cannot use $q$ which is taken already for $q=Q/2|\gamma|^{1/2}$).

 In a static spherically-symmetric spacetime the motion is planar and without loss of generality the orbital plane can be taken as $\theta = \pi /2$. The equations of motion may be reduced to the two following first order equations which we write for the dimensionless radial coordinate $x(\tau)=r(\tau)/\ell_{el}$ and the azimuthal angle $\phi(\tau)$ as:
\be
x^2 \dot{\phi} = \ell \,\,\, ; \,\,\,\,\,\, \dot{x}^2+f(x)\left( \epsilon +\frac{\ell^2}{x^2} \right) - \frac{({\cal E}-q_{_{2}}v(x))^2}{a^2(x)} = 0
\label{GeodEqsElectricIntegrated}\ee
where $q_2$ is the dimensionless product of the intercating charges $Q$ and $q'$ and $v(x)$ is the dimensionless EHBH potential.

In order to understand the motion of test particles around EHBHs, we will plot for a couple of representative BHs the effective potential which can be read off directly from Eq. (\ref{GeodEqsElectricIntegrated}), although the specific metric functions have to be substituted numerically. We will concentrate on neutral particles only. Then we will turn to light rays and light deflection around EHBHs.
\begin{figure}[b!!]
\begin{center}
{\includegraphics[width=8cm, angle = -00]{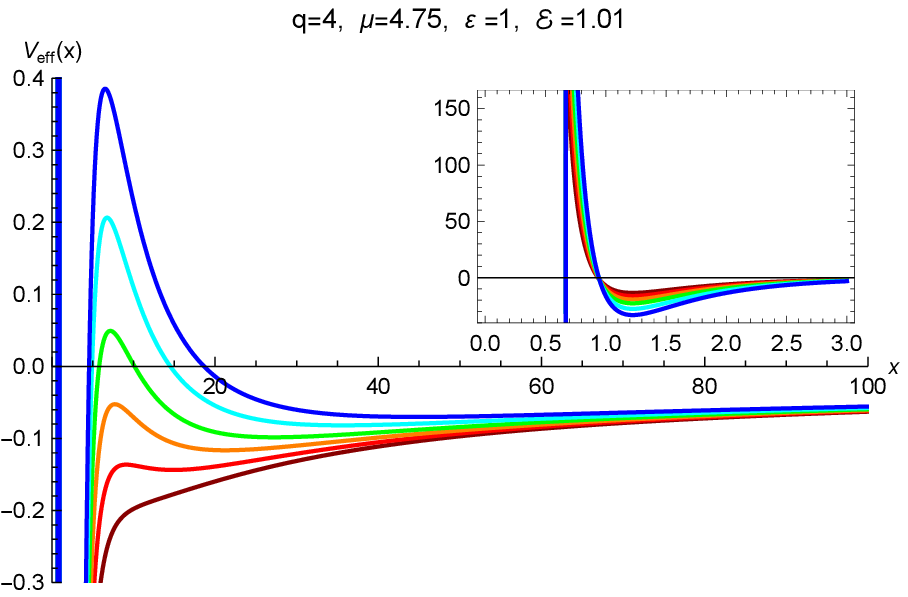}}
{\includegraphics[width=8.25cm, angle = -00]{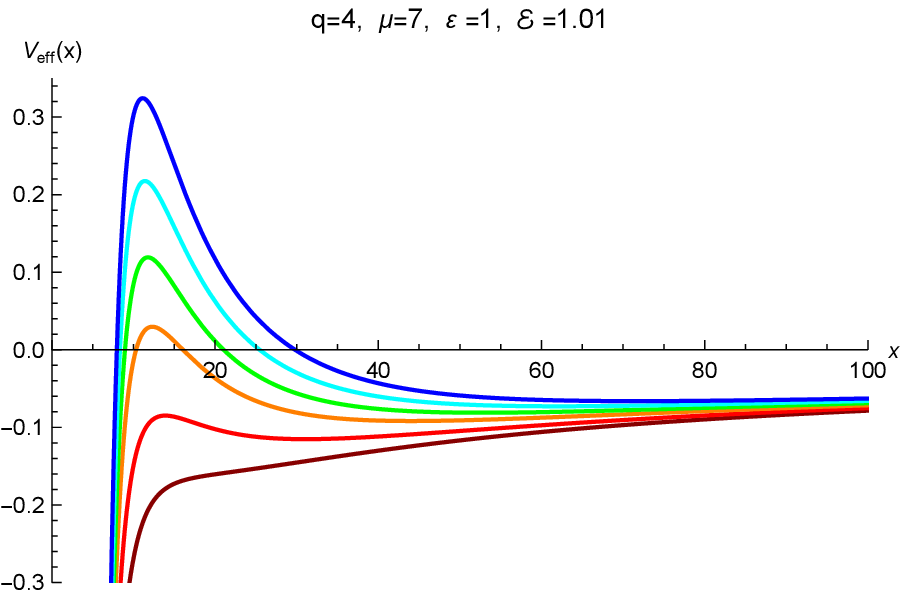}}
\end{center}
\caption{{\small The effective potential for particle trajectories with  rescaled particle energy of ${\cal E}=1.01$ around two of the $\gamma>0$ electric HBH solutions presented in Fig. \ref{FigProfilesElectricGamPos}. Left: RN-like EHBH with $\mu=4.75$, curvature singularity at $x_s = 1.4939$ and event horizon at $x_H=3.6635$. The curves correspond to the following values of rescaled angular momentum: $\ell = 6.75,\; 7.5,\; 8.25,\; 9,\; 10,\; 11$. The insert shows a ``zoom-out'' of the vicinity of the curvature singularity where $V_{eff} (x)$ diverges to $-\infty$. Notice that this domain lies within the event horizon. Right: S-like EHBH with $\mu=7$, and event horizon at $x_H=6.3734$. The $\ell$-values are: $\ell = 11,\; 12.5,\; 14,\; 15,\; 16,\; 17$.}} \label{FigElBHVeffGamPos}
\end{figure}
Fig. \ref{FigElBHVeffGamPos} presents the main features of the effective potential by choosing two EHBHs - one of each of the two types with $\gamma>0$: S-like and RN-like which possesses a spherical surface of curvature singularity. Unlike Fig. \ref{VeffTimelike}, we present here only the curves for open trajectories where ${\cal E }>1$. For each BH the effective potential curves  are demonstrated for several $\ell$ values and we notice that they exhibit outside the horizon a similar structure to that we met in the magnetic case. This is not surprising in light of the similarity (outside the horizon) of the metric components in MHBHs and EHBHs of the same $\mu$ parameter and $q=p$. The structure inside the horizon is significantly different in both cases as is evident from looking at Figs. \ref{FigElBHVeffGamPos} and \ref{VeffTimelike}. There are some subtleties of the effective potential inside the horizon, but we will not discuss them here.

We turn now to the issue of light deflection. Although we do not have the explicit EHBH metric components, it is still possible to get an analytic expression in the form of an expansion in powers of the inverse periastron distance. The expansion up to third order for the deflection angle for light is completely analogous to that in the magnetic case of (\ref{DeflectionExpansion}) with the only replacement $p \mapsto q$:
\be
\psi(\mu,q,s_p)=2\mu s_p +\left[\left(\frac{15 \pi
   }{16}-1\right) \mu ^2-\frac{3 \pi  q}{4}\right] s_p^2+
   \left[\left(\frac{61}{12}-\frac{15 \pi }{16}\right) \mu ^3-\left(7-\frac{3 \pi }{4}\right) \mu  q\right] s_p^3 +\cdots
\label{DeflectionExpansionElectric}\ee
The parameter $\mu$ is the dimensionless mass parameter which is related to the asymptotic value of the mass function $m(x)$ which we used in this and previous section by $\mu = q m(\infty)$. As in the magnetic case, there is no dependence on the sign of $\gamma$ up to third order.

Notwithstanding the identical power series of the light deflection angle in the electric and magnetic case, this phenomenon does not extends to the full domain. Actually, the similarity breaks already at the fourth order of the expansion of the deflection angle. A similar difference is also seen in the asymptotic expansion of the effective potential magnetic vs the electric HBHs. Still, quantitatively the differences are usually very small in many circumstances as seen for instance by a comparison of the numerically obtained  $\psi(\mu,q,s_p)$ for the electric case  with Eq.(\ref{DeflectionGeneral}) for the magnetic case. For example, the analogous figure to the left panel of Fig.\ref{FigLightDeflection-MagFixed-p} for the electric case (i.e. a EHBH with $q=4$) looks very almost identical, with the only difference that the mass parameter of the extremal EHBH with $q=4$ is a little smaller and is $\mu=3.962$, so the corresponding curve is different from that of the $\mu=3.988$ of the extremal MHBH. However, the $\psi(\mu,q,s_p)$ curve of the non-extremal EHBH with $\mu=3.988$ is almost identical to that of the extremal  MHBH with the same $\mu$ value. On the other hand, we notice that the right panel of Fig.\ref{FigLightDeflection-MagFixed-p} has no electric analog. In short, in most cases where for a given $\mu$ there exist HBHs with $q=p$, they both produce very similar light deflections. Therefore, there is no need to add new plots of $\psi(\mu,q,s_p)$ for the electric case.  This near symmetry between the electric and magnetic cases may be considered somewhat surprising in a theory which lacks the duality symmetry. However, as mentioned already it is broken at the order of $s_p^4$ in $\psi(\mu,q,s_p)$ as is of course expected from the difference between the asymptotic expansions of $f(r)$ in the the two cases. The symmetry between positive and negative $\gamma$ is also broken at the same order.

\section{Conclusion}\label{conclusion}
    In this paper we constructed the magnetically charged black hole solution of the Einstein-Maxwell system with an additional non-minimal coupling between the vector field and the gravitational field of the form
    $^{**}R_{\kappa \lambda}^{\phantom{\kappa \lambda} \mu \nu} F^{\kappa \lambda}F_{\mu \nu}\sim\,^*F^{\kappa\lambda}\,\, ^*F^{\mu\nu}R_{\kappa\lambda\mu\nu}$,
    and performed a detailed comparison to their electric counterparts. Unlike the analogous electric case, the field equations become linear  and they  can be  solved analytically in terms of hypergeometric functions and other elementary functions. These solutions describe new kinds of black holes. Some of the solutions are similar to the ordinary RN solutions, but others have various exotic properties. Some may have repulsive gravitational field around them, while others have a spherical curvature singularity rather than a point-like. We have calculated their mass-charge-horizon relations, their temperature behavior and other characteristics. We found a significant difference between positive and negative $\gamma$. For $\gamma>0$ all BHs are RN-like, i.e. $g_{00}(r)\rightarrow \infty$ as $r\rightarrow 0$. On the other hand, for $\gamma<0$ all BH solutions have a spherical surface of curvature singularity but they are divided into two subclasses: RN-like and S-like. This difference is reflected by their temperature dependence on their masses: The temperature of an RN-like BH starts at zero in an extremal solution, increases with $M$ until it reaches its maximum, and then decreases. The S-like BHs have a temperature with similar decreasing behavior for large $M$, but the temperature is not bounded from above but diverges for a minimal (non-zero) value of the mass given by Eq. (\ref{S-like-Mass-Minimum}). We also studied extensively the trajectories of point particles around these BHs - charged, neutral and photons. Next, we moved to the electrically charged BHs in order to compare them to the magnetic ones. First we were able to reduce the field equations to a very simple universal system (without free parameters) of two decoupled first order equations for the dimensionless mass function $m(x)$ and the metric function $a(x)$. Although this system still cannot be solved analytically, it is simple enough that plenty of general conclusion can be drawn directly and intuitively. Yet, the field equations can be readily solved (numerically) thus enabling us to identify again the existence of S-like and RN-like solutions and the domains in parameter space where each of them exist. Incidentally, we could show that there are no negative mass EHBHs unlike the magnetic case. Consequently, we were able to obtain the relation between the BH mass, charge and horizon size, the properties of the temperature function $T(\gamma, Q, M)$ and compare systematically the electric with the magnetic BHs. As done for the MHBHs, we analyzed particle trajectories around the EHBHs and light deflection. We calculated the deflection angle in both cases as a power series in $1/r_p$ up to third order and found them to be identical. We found further that differences start to appear in fourth order.

    Still, further analysis of these systems is in order. Some of the immediate directions are: studying more systematically the negative mass MHBHs for $\gamma>0$ and the properties of the spherical singularity of the BH solutions for $\gamma<0$, studying the thermodynamics of the magnetic \cite{Feng+Lu2015} and electric BHs, generalizing to the non-Abelian case and coupling scalar fields and discussing scalarization. Another direction of study is the question whether the fact that the field equations become linear in the magnetic case enables to superpose these elementary solutions to construct multi-center solutions and other richer structures.
\\
\\
\noindent {\bf Acknowledgments:} It is a pleasure to thank G. Horndeski for a helpful correspondence and valuable comments.

\section{Appendix: Solving Equation (\ref{Eqfz})}
\setcounter{equation}{0}

The equation (\ref{Eqfz}) is:
\be
\label{EqfzAgain}
 4(z+1)z f' +(6z-1)f  -  p\, z^{1/2} + 1=0
\ee
This is a linear inhomogeneous ODE. The standard procedure to solve this equation is to obtain the general solution of the homogeneous equation. This is a trivial matter of a direct integration:
\be
\label{HomSol}
 f_{0}(z)=c_{1}\frac{z^{1/4}}{ (1+z)^{7/4}}
\ee
Next we replace the integration constant $c_{1}$ by a function $u(z)$, so the solution is the product $ f(z)= f_{0}(z) u(z)$. Consequently, the function $u(z)$ must solve
\be
\label{EqForSpecialSol}
 4(z+1)z f_{0}(z) u'  -  p\, z^{1/2} + 1=0
\ee
The solution for $u(z)$ is again obtained by direct integration which is expressed in terms of two hypergeometric functions:
\be
\label{SpecialSol}
 u(z) = p z^{1/4}F\left(-\frac{3}{4},\frac{1}{4},\frac{5}{4},-z\right) + \frac{1}{z^{1/4}}F\left(-\frac{3}{4},-\frac{1}{4},\frac{3}{4},-z\right) - \mu
\ee
where $\mu$ is the integration constant. The integration is performed by the following identity for hypergeometric functions:
\be
\label{HyperGeomIdent}
\int  (1+z)^\alpha z^\beta dz= \frac{z^{\beta+1}}{\beta+1}F\left(-\alpha,\beta+1,\beta+2,-z\right)
\ee
which is obtained in a straightforward way from the standard properties of the hypergeometric functions (see e.g. Abramovitz and Stegun \cite{A+S}  ). The function $f(z)$ which solves Eq. (\ref{Eqfz}) is just a product of the two functions in (\ref{HomSol}) and (\ref{SpecialSol}):
\be
\label{FinalSolToEq-f}
f(z)= \frac{z^{1/4}}{ (1+z)^{7/4}} \left[   p z^{1/4}F\left(-\frac{3}{4},\frac{1}{4},\frac{5}{4},-z\right) + \frac{1}{z^{1/4}}F\left(-\frac{3}{4},-\frac{1}{4},\frac{3}{4},-z\right) - \mu   \right]
\ee

\vskip 0.5cm



\end{document}